\documentclass[11pt]{article}

\usepackage[letterpaper,margin=1in]{geometry}
\usepackage[T1]{fontenc}
\usepackage[utf8]{inputenc}
\usepackage{setspace}
\usepackage{times}
\usepackage{microtype}
\usepackage{amsmath,amssymb,amsfonts,amsthm}
\usepackage{mathtools}
\usepackage{bm}
\usepackage{graphicx}
\usepackage{booktabs}
\usepackage{caption}
\usepackage{subcaption}
\usepackage{float}
\usepackage{enumitem}
\usepackage{authblk}
\usepackage{xcolor}
\usepackage{natbib}
\usepackage{hyperref}

\hypersetup{
    colorlinks=true,
    linkcolor=blue,
    citecolor=blue,
    urlcolor=blue
}

\theoremstyle{definition}
\newtheorem{hypothesis}{Hypothesis}

\newif\ifanon
\anonfalse  

\title{\bfseries Trading in the Sunshine or in the Shade:\\ Market Impact, Liquidity, and Adverse Selection on Hyperliquid
}

\ifanon\author{}\date{}\else
\author[1]{Davide Barone}
\author[1]{Fabrizio Lillo}
\affil[1]{Scuola Normale Superiore, Pisa, Italy

\texttt{davide.barone@sns.it; fabrizio.lillo@sns.it}}

\date{2nd INFORMS Conference on Financial Engineering and FinTech\\
November 6--7, 2026, Stevens Institute of Technology}

\fi

\begin{document}
\maketitle

\begin{abstract}
\noindent Whether a market should let large traders disclose their intentions before they trade is a first-order question in market design. Theory is divided: preannouncement can separate uninformed liquidity demand from informed order flow and attract liquidity provision, lowering execution costs, or it can invite front-running and predation. Direct evidence is scarce because credible preannouncement of large orders is almost never observed. We provide such evidence using Hyperliquid, a fully on-chain limit order book  for cryptocurrency perpetual futures. In this market a large order can be executed as a time-weighted average price (TWAP) order whose terms are public from inception, constituting a natural form of sunshine trading. We compare 465{,}000 of these preannounced orders with 4.3 million large orders that traders split into many trades without announcing them, reconstructed from account-level data. We test four predictions of \citet{AdmatiPfleiderer1991}. We find that announced orders move the price by about 10 basis points less during execution and leave roughly 5 basis points less permanent price impact than comparable unannounced orders. We detect no cost shifted onto traders who do not announce. The displayed order book tilts toward the side that absorbs an announced order, more so for large orders, and other traders trade less in its direction. Finally, we find little to no evidence of front-running of preannounced orders.
\end{abstract}

\vspace{0.5em}

\vspace{0.5em}

\section{Introduction}

How much of their trading intentions should market participants be allowed---or required---to reveal before they trade? Pre-trade transparency is one of the central levers of market design, and both exchanges and regulators must repeatedly decide how much pending trading interest to display and to whom \citep{MadhavanPorterWeaver2005,BoehmerSaarYu2005}. The decision is consequential precisely because its welfare effects are theoretically ambiguous. On one side, an influential argument holds that a large trader who can credibly announce that an order is \emph{uninformed} separates that order from the potentially informed flow, mitigates the adverse selection that market makers would otherwise price in, and can attract additional liquidity supply---so that transparency lowers execution costs \citep{AdmatiPfleiderer1991}. On the other side, disclosure exposes the trader to anticipation and predation: others may front-run the announced order and move prices against it \citep{BrunnermeierPedersen2005}. Whether those who learn of the order prey on it or supply liquidity to it depends on the liquidity of the market and on how many they are, and so does the choice between announcing and trading in secret \citep{schoeneborn_schied_2009}. Which force dominates is ultimately an empirical question, and one of direct relevance to how trading venues are designed.

The empirical difficulty is that credible preannouncement of large orders is almost never observed. A large trading decision reaches the market progressively. The decision itself is the \emph{parent order}: the instruction to buy or sell a given total quantity of an asset within a given horizon, as the investor hands it to a trading desk or to an execution algorithm. It is then worked through many smaller \emph{child orders} \citep{BattalioHatchSaglam2024}. Institutions with large orders routinely split them in this way \citep{lillo_farmer_2004_long_memory,toth_palit_lillo_farmer_2015_persistent}, whether to conceal private information \citep{kyle1985continuous} or to contain execution costs \citep{BertsimasLo1998,almgren2001optimal}. The sequence of child orders that executes a parent order is a \emph{metaorder}. The two are distinct objects: the parent order is an intention, normally known only to the trader, while the metaorder is what the market sees, one child at a time, without being told that the trades belong together. In this paper we call an order \emph{visible} when the parent order itself is public before its children execute, and \emph{hidden} when only the children are. Visible execution is our counterpart of sunshine trading, the preannouncement of an order studied by \citet{AdmatiPfleiderer1991}; Section~\ref{sec:setting} gives the definitions we use in the data. Genuine sunshine trading is rare, so the counterfactual---what the same order would have cost had it been hidden---is unobserved, and the available evidence is indirect: it rests on episodes such as the temporary display of marketable orders through Nasdaq's flash functionality \citep{SkjeltorpSojliTham2012} or the predictable futures rolls of a large commodity exchange-traded fund \citep{BessembinderCarrionTuttleVenkataraman2016}. These studies are broadly consistent with transparency improving market quality, but they do not observe, side by side, the cost of executing a parent order visibly and the cost of executing it in the shade.

This paper exploits a setting that supplies exactly this comparison. Hyperliquid is a fully on-chain central limit order book (CLOB) for cryptocurrency spot and perpetual-futures markets on which two features coincide. First, the protocol offers a native time-weighted-average-price (TWAP) order that automatically slices a parent order into regularly spaced child orders; crucially, the parent order's direction, size, and horizon become publicly visible the moment it is submitted and remain visible while it executes. A native TWAP is therefore a credible, protocol-level preannouncement of trading intent---an institutional analogue of sunshine trading embedded in an otherwise standard electronic market. Second, because every action on the venue is attributable to an on-chain address, we can reconstruct the hidden benchmark: sequences of same-address, same-direction market orders that constitute hidden metaorders, whose parent order is never observed. The comparison is thus not between two trading algorithms, but between two \emph{ways of executing} a large order that differ in ex-ante observability, which we compare on a common support of size, duration, and participation.

We take the four predictions of the sunshine-trading model of \citet{AdmatiPfleiderer1991} to this setting, with 4.3 million hidden metaorders and 465{,}000 visible metaorders (native TWAPs) across 199 perpetual markets. The first two concern the trader who announces. \emph{(i)~Announcers pay less}: at the median volatility and conditional on execution characteristics, visible metaorders incur about $10$ basis points lower temporary impact than comparable hidden metaorders. \emph{(ii)~Announced flow leaves less permanent impact}: about $5$ basis points less than hidden flow after completion, consistent with lower information content. The two groups keep a similar share of their impact at completion, so what is lower is the level of the permanent impact, not the share of the move that reverts. These comparisons are observational: traders choose whether to announce, and most use only one of the two execution modes. The traders who use both allow a within-trader comparison, and the same trader pays less and leaves about 7 basis points less permanent price impact when announcing than when executing in the shade.

The other two predictions concern everyone else. \emph{(iii)~Nonannouncers do not pay detectably more}: here the prediction depends on the mechanism. With private information, announcing makes the flow that stays hidden look more informed and raises its cost, by an amount that grows with the share of liquidity demand that is announced, which is small in this market; with costly entry of liquidity suppliers, the cost can rise or fall. In the data, the cost of hidden metaorders does not change detectably when a market's announced share rises. Hidden metaorders do pay more when other metaorders trade alongside them in the same direction, but as much when these are hidden as when they are visible: this is the price pressure of the flow, not a cost of its announcement. \emph{(iv)~The book tilts toward the announced order}: while a visible metaorder executes, the displayed book tilts toward the side that must absorb it, more so for larger long programs, and other traders trade less in the announced direction.  We find little to no evidence of front-running of the announced order.

Along the way we document that the two groups also differ sharply in execution style, and consequently in the price trajectory during execution: hidden metaorders follow U-shaped schedules and double-inflection price paths, consistent with strategic execution under transient price impact, whereas visible metaorders, which trade uniformly by construction, trace a standard concave price trajectory.

Our contribution is threefold. First, we provide, to the best of our knowledge, the first direct test of the classic predictions of sunshine trading in a live, continuous electronic limit order book, using observable preannouncement together with a reconstructed hidden benchmark rather than indirect evidence. The evidence favors the liquidity-provision reading of transparency: on average, visible flow is less costly, less informationally adverse, and met by a book that tilts toward it rather than by predation. Second, we treat transparency as a question of market design. As on-chain order books come to compete with centralized exchanges, transparency ceases to be only a trader's disclosure choice and becomes a protocol-level design variable, to be weighed against the well-documented risks of on-chain order-flow exposure such as front-running and maximal extractable value \citep{daian2020flashboys,qin2022quantifying_bev}. Our evidence concerns voluntary disclosure. Hyperliquid lets a trader announce an order but does not require it, so we observe the orders that traders choose to reveal; what would happen to prices and liquidity if disclosure were mandatory, our results do not say. Third, our measurements speak to practice: they quantify the execution-cost advantage of visible scheduled execution, document what hidden traders pay while visible programs are active, and show how displayed liquidity responds to announced demand---inputs to the design of execution algorithms, of venue transparency rules, and of the disclosure policies that regulators debate.

The remainder of the paper is organized as follows. Section~\ref{sec:literature} positions the paper in the literature. Section~\ref{sec:hypotheses} develops the sunshine-trading framework and states the four hypotheses. Section~\ref{sec:setting} describes the Hyperliquid setting, the construction of the visible and hidden samples, and descriptive statistics. Section~\ref{sec:results} presents the results: it first compares the aggregate cost and the execution dynamics of visible and hidden metaorders (Section~\ref{sec:regimes}) and then tests the four hypotheses (Section~\ref{sec:tests}). Section~\ref{sec:robustness} addresses identification and robustness. Section~\ref{sec:conclusion} discusses implications for market design and policy and concludes. The appendices report additional results (Appendix~\ref{app:additional}), the estimates by trader composition (Appendix~\ref{app:traders}), alternative constructions of hidden metaorders (Appendix~\ref{app:construction}), and visible metaorders that stop before their announced size is filled (Appendix~\ref{app:interrupted}).

\section{Related Literature}
\label{sec:literature}

Our work connects three strands of research: the market-design literature on pre-trade transparency and sunshine trading, the microstructure literature on the price impact and cost of large orders, and the emerging literature on the quality of decentralized markets.

\paragraph{Pre-trade transparency and sunshine trading.}
The theoretical benchmark is \citet{AdmatiPfleiderer1991}, who show that allowing a trader to preannounce order flow can, by revealing that the flow is uninformed, lower the announcer's expected cost, raise the cost borne by nonannouncers in the baseline model, and make prices more informative; a companion mechanism with costly entry lets liquidity suppliers condition participation on announced demand. \citet{Madhavan1996} studies the case in which the market discloses the order flow of liquidity traders to everyone: transparency can then lower liquidity and raise volatility, although in a sufficiently large market it improves market quality. \citet{deFrutosManzano2014} make the decision to preannounce endogenous, with strategic informed traders: whether a liquidity trader announces does not depend on the size of the order, and traders with larger liquidity needs gain more from announcing. The idea that order flow identified as uninformed trades at better terms is older than sunshine trading: it underlies the analysis of dual-capacity trading in \citet{Roell1990}, where costs fall for liquidity traders who can convince a dealer that they are uninformed and rise for those who cannot, of anonymity in \citet{ForsterGeorge1992}, and of block trades negotiated upstairs, where the broker certifies that a trade is not information-motivated \citep{Seppi1990,MadhavanCheng1997,BessembinderVenkataraman2004} and draws on liquidity that is not displayed in the book \citep{Grossman1992}. The countervailing view stresses that disclosure can trigger anticipation and predatory trading, so that its net effect depends on market structure and the balance between adverse selection and strategic exploitation \citep{Harris1997,BrunnermeierPedersen2005,CarlinLoboViswanathan2007,schoeneborn_schied_2009}. \citet{Harris1997} states the trade-off from the trader's side: traders expose their orders to encourage others to trade with them, and avoid exposing when they fear front-runners and quote-matchers.

Empirically, transparency has been studied largely through changes in the visibility of the limit order book or through episodes of selective disclosure. \citet{BoehmerSaarYu2005} find that the NYSE's OpenBook reform, which revealed depth to off-floor traders, altered order strategies and improved informational efficiency and liquidity; \citet{MadhavanPorterWeaver2005} find instead that execution costs and volatility rose when the Toronto Stock Exchange began to display its limit order book, and theory, experiments and market reforms show that the effect depends on what is displayed and on who is identified \citep{PaganoRoell1996,BloomfieldOHara1999,HendershottJones2005,FoucaultMoinasTheissen2007,BloomfieldOHaraSaar2015}. Closest to us, \citet{SkjeltorpSojliTham2012} use Nasdaq flash orders as a form of sunshine trading and find that selective disclosure improved liquidity, and \citet{BessembinderCarrionTuttleVenkataraman2016} show that liquidity and resiliency improve around the predictable futures rolls of a large commodity ETF. \citet{DiaPouget2011} document order placement consistent with sunshine trading in the pre-opening of a call auction on the West African Bourse. We add to this literature a setting with \emph{explicit, credible, order-level} preannouncement and an observable hidden benchmark, which lets us test the announcer, nonannouncer, and liquidity-supply predictions jointly rather than through aggregate market-quality changes.

\paragraph{Metaorders, price impact, and execution costs.}
Measuring the cost of executing a large order is the object of a long microstructure and execution literature. Since \citet{kyle1985continuous}, informed traders are understood to split large orders to balance execution cost against information leakage, and optimal-execution models trade impact against timing risk \citep{BertsimasLo1998,almgren2001optimal} and exploit the resilience of liquidity after each trade \citep{ObizhaevaWang2013}. Empirically, the temporary impact of a metaorder scales as a concave, approximately square-root function of size relative to volume \citep{almgren2005direct,toth2011anomalous,donier2015million,zarinelli2015_beyond_square_root}, and impact has a substantial transient component that partially reverts after completion, leaving a smaller permanent price impact tied to information \citep{moro2009market,BacryIugaLasnierLehalle2014,waelbroeck2013market}. We use these regularities as the measurement toolkit for execution cost---temporary impact, its decomposition into participation and duration, price trajectories, and post-trade decay---rather than as objects of interest in themselves, and we interpret the differences between visible and hidden metaorders through the economics of transparency. A related literature asks how intermediaries respond to large orders that are \emph{not} announced: high-frequency traders initially lean against large institutional orders but trade with them once they persist \citep{vanKervelMenkveld2019}; they submit more orders in the direction of a large institutional trade while it executes, and their activity lowers the cost of small, uninformed trades but raises that of large, informed ones \citep{KorajczykMurphy2019}; and aggressive fast traders anticipate the order flow of others \citep{Hirschey2021}, a behavior modeled by \citet{YangZhu2020}. Exposure to a few counterparties can be costly: when the pieces of an institutional order are routed to electronic liquidity providers, each piece trades at a better price but the order as a whole costs more, because the providers detect the parent order quickly \citep{BattalioHatchSaglam2024}. Because these regularities are documented in \emph{anonymous} markets, where traders hide intentions, they also frame our central question: how execution cost and its dynamics change when intentions are instead revealed ex ante.

\paragraph{Decentralized markets, transparency, and market quality.}
Blockchain-based venues make order flow and execution outcomes publicly observable, turning transparency into a protocol-level design variable. A growing literature compares the market quality of centralized and decentralized venues \citep{BarbonRanaldo2026,LeharParlour2025,CapponiJiaYu2026} and documents the risks that on-chain visibility creates---front-running, transaction-reordering, and maximal extractable value \citep{daian2020flashboys,zhou2021hft_dex,qin2022quantifying_bev,Park2023}---which motivate private mempools, batch auctions, and related protections. On-chain CLOBs such as Hyperliquid are more directly comparable to classical electronic markets than automated market makers, and recent work uses Hyperliquid's message data to study its microstructure \citep{albers2026openbook,albers2026neutrinos}. We use this transparency constructively: rather than measuring the cost of exposure, we ask whether a protocol-level form of preannouncement delivers the benefits that sunshine-trading theory predicts, and thereby connect the market-design question to the practical design of decentralized trading systems.

\section{Methodology}
\label{sec:hypotheses}

Our design compares two groups of large orders---\emph{visible} protocol-native TWAP orders, whose direction, size, and horizon are publicly disclosed at inception, and \emph{hidden} metaorders, reconstructed ex post from address-level flow---which we define in detail in Section~\ref{sec:setting}. This section develops the theoretical predictions we take to that comparison.

We organize the empirical analysis around the sunshine-trading framework of \citet{AdmatiPfleiderer1991}, which separates two effects of preannouncement. The first is informational. In their baseline model with heterogeneous private information and zero entry costs, all speculators are already present, so preannouncement does not change the size of the market; it changes what the market infers from order flow. When announced demand is identified as coming from liquidity traders, it is separated from potentially informed flow. Preannouncement then lowers the expected trading cost of announcers, raises the expected cost of nonannouncers, lowers the aggregate cost of liquidity traders, and makes prices more informative. The second effect operates through liquidity supply. In a model with positive entry costs, speculators pay a fixed cost to enter; preannouncement lets entry condition on the realized announced demand, which can raise the effective supply of liquidity around announced orders when entry costs are high, though the effect can reverse when entry is cheap. With costly entry the effect of preannouncement on the traders who do not announce is ambiguous.

We call native TWAP executions visible metaorders and statistically reconstructed ones hidden metaorders; the first are the empirical counterpart of announced trades in \citet{AdmatiPfleiderer1991}, the second of unannounced ones. \emph{Hidden} refers to a parent order that is never disclosed, not to non-displayed limit orders.  The correspondence is in disclosure, not in information types. In the model, announcers and nonannouncers are both uninformed liquidity traders, distinct from speculators with private information. In our data, trading motives are unobserved: traders executing visible metaorders need not be uninformed, and hidden metaorders may reflect either liquidity needs or private information.  The comparisons below consequently concern the costs and price impact of announced and unannounced execution, rather than the effect of disclosure holding traders' information fixed. Three features of the setting shape interpretation and are made explicit here. First, traders \emph{self-select} into visible execution, so the uninformed nature of announced flow is not imposed by the protocol but is a property to be tested. \citet[Section~4.1]{AdmatiPfleiderer1991} argue that it is a natural one: announcing commits the trader to a slow and public execution, which is more costly for a trader whose information is short-lived. The comparison is therefore between visible and hidden metaorders as traders choose them, rather than a randomized manipulation of a single order's visibility (we return to this in Section~\ref{sec:robustness}). Second, a hidden metaorder is not anonymous: address-level flow can be statistically inferred as it unfolds, a distinction that matters for the book-level test below. Third, all four hypotheses presume that the announcement is observed: a native TWAP is public from the block that records the order, and its first child order executes between one and two seconds later. Section~\ref{sec:setting} describes what market participants can observe and when.

Before testing the four transparency predictions, a preliminary comparison is warranted. Visible and hidden metaorders differ not only in whether they are announced but also in how they are executed, and a difference in cost could come from either. We therefore first compare their execution style and aggregate cost (Section~\ref{sec:regimes}), and then test the four hypotheses that the transparency mechanism implies, which we now state.

\begin{hypothesis}[Announcer costs]\label{hyp:announcer}
Visible metaorders incur lower execution costs than comparable hidden metaorders, as predicted for announcers in the informational model and in the costly-entry model when the entry margin is important.
\end{hypothesis}

\begin{hypothesis}[Information content]\label{hyp:information}
Visible flow is on average less informationally adverse than hidden flow, leaving a smaller permanent price impact after completion.
\end{hypothesis}

\begin{hypothesis}[Nonannouncer costs]\label{hyp:nonannouncer}
The effect of preannouncement on nonannouncers' trading costs is ambiguous. In the informational model, identifying announced liquidity demand as uninformed raises the adverse-selection costs borne by liquidity traders who do not announce. In the second model, with costly entry by speculators and no private information, preannouncement allows liquidity suppliers to condition entry on announced demand and can lower nonannouncers' trading costs when entry costs are sufficiently high.
\end{hypothesis}

\begin{hypothesis}[Liquidity provision]\label{hyp:liquidity}
The limit order book supplies liquidity around visible execution: during active windows of visible metaorders the book tilts toward the absorbing side, and the liquidity displayed there increases with size, as predicted by the costly-entry model, in which entry is increasing in the size of the announced order.
\end{hypothesis}

A related conjecture concerns liquidity demand. \citet[Section~4.7]{AdmatiPfleiderer1991} note that liquidity traders with discretion over timing may trade against a preannounced order, which further lowers the announcer's cost. We examine it together with Hypothesis~\ref{hyp:liquidity}, by asking whether other traders take more liquidity against the announced direction, and less in it, while a visible metaorder is active.

\section{Data and Empirical Design}
\label{sec:setting}

Our analysis uses publicly observable data from Hyperliquid, a purpose-built Layer-1 blockchain that hosts fully on-chain central limit order books for spot and perpetual-futures markets.\footnote{Hyperliquid documentation, \url{https://hyperliquid.gitbook.io/hyperliquid-docs} (accessed June 1, 2026), which is the source of the protocol rules described in this section.} Its consensus protocol, inspired by HotStuff \citep{hotstuff_podc2019}, is optimized for latency: according to its documentation, for a co-located client end-to-end latency has a median of $0.2$ seconds and a $99$th percentile of $0.9$ seconds. In our data, the median interval between consecutive blocks is \(0.08\) seconds (5th and 95th percentiles: \(0.05\) and \(0.13\)). The order books are part of the state of the chain. Every order, cancellation and modification, including those that are rejected, is a transaction that consensus orders and records in a block, and trades result from executing these transactions in that order. The activity of each address is therefore public message by message. An address is the identifier of an account on the chain: it is public and persistent, but it does not reveal who owns it, and one trader can hold several.

We study 199 Hyperliquid perpetual markets from July~28, 2025 to March~23, 2026. These instruments are crypto perpetual futures: linear contracts on the oracle index of the underlying spot asset, with no expiration and hourly funding. Perpetuals are USDC-margined and USDT-denominated; we measure order size, market volume, and traded value as USDC notional, which we refer to as USD since USDC is pegged one-to-one to the dollar. We use three sets of data. Trades come from Hydromancer's Reservoir, a public Hyperliquid data layer,\footnote{Hydromancer, Reservoir: Hyperliquid Historical Data, \url{https://docs.hydromancer.xyz/reservoir} (accessed May 27, 2026).} which records every fill: one row for each side of a trade, with its time, market, direction, price and size, the address and the order it belongs to, whether that order took or provided liquidity, and flags for liquidations and for the child orders of native TWAPs. We reconstruct metaorders from these fills. The sample contains more than 641 million fills, corresponding to about 365 million market orders and roughly USD~1.93~trillion in traded volume. About \(550{,}000\) addresses trade in these markets over the sample, and trading is concentrated: \(1\%\) of them execute \(86\%\) of the volume of market orders. The order book comes from 0xArchive:\footnote{0xArchive, \url{https://0xarchive.io} (accessed November 16, 2026).} \(6.3\) billion snapshots, about two per second, with the price and size of the ten best levels on each side. We take every midquote and every book measure from these snapshots. We checked both sources against the data that Hyperliquid publishes from its own nodes, on days and hours drawn at random: fills and snapshots are identical.\footnote{On six days drawn at random among those for which the nodes' data are available, the \(13.4\) million fills of our markets in Hydromancer are identical, fill by fill, to those recorded by the nodes, including the addresses of both counterparties, and no fill recorded by the nodes is missing. On \(2{,}244\) market-hours (twelve hours drawn at random, 187 markets), the \(13.6\) million 0xArchive snapshots are identical to the nodes' snapshots of the same instant, up to the tenth price level.} Finally, for each native TWAP we use the on-chain record of the parent order: the time of the announcement, the planned size and duration, its options, and how it ended. To group addresses into traders we use the role that the protocol reports for each address: sub-accounts are assigned to their master account, and vaults, accounts that trade the pooled funds of their depositors, are identified as such. Because markets trade 24 hours a day, seven days a week, we compute normalization variables such as daily volume and volatility over a 24-hour forward-looking window anchored at each metaorder's start. Measuring them over the 24 hours before the start gives about the same estimates (Appendix~\ref{app:robustness}).

\subsection{Visible and \texorpdfstring{Hidden}{Hidden} \texorpdfstring{Metaorders}{Metaorders}}
\label{sec:regimes-definition}

The key feature of the data is wallet-level attribution: each action is assigned to the initiating on-chain address. We identify metaorders in two ways, corresponding to the two groups we compare.

\paragraph{Visible execution: protocol-native TWAP orders.} Hyperliquid offers native TWAP execution as a protocol-level mechanism that splits a metaorder into child market orders submitted automatically every $30$ seconds over a running time chosen by the trader, between five minutes and 24 hours. The protocol keeps execution close to a time-proportional schedule subject to a maximum per-slice slippage of $3\%$, and allows later slices to grow when earlier ones underfill, with catch-up bounded at three times the normal suborder size. During our sample the trader could also randomize the size of each suborder by up to \(\pm20\%\) (\(29\%\) of TWAPs in the on-chain order records) and make the order reduce-only (\(12\%\)). Other options, such as trigger prices and price limits, were introduced in August 2026, after the end of the sample.\footnote{Hyperliquid announcement ``TWAP order improvements'', August 2, 2026, \url{https://app.hyperliquid.xyz/announcement/ih9j0po0zx}.} Of the TWAPs of our markets that trade at least once, \(51\%\) run to completion, \(34\%\) are cancelled by the trader and \(15\%\) are stopped by the protocol, almost always for insufficient margin. We keep them in our sample whatever their outcome: the main estimates are consistent with those obtained from the TWAPs that run to completion alone (Appendix~\ref{app:robustness}), and Appendix~\ref{app:interrupted} studies those that stop early. 

These orders are the empirical counterpart of sunshine trading for two reasons. First, the full parent instruction---direction, total size, execution horizon, and other parameters---becomes publicly visible as soon as the order is recorded in a block, so other participants can observe the program before it completes. The progress of the program is public as well: every child fill is a public trade carrying the TWAP identifier, and a cancellation is itself an on-chain action, visible in the block that records it.\footnote{Active TWAPs can be inspected, for example, on HypurrScan: \url{https://hypurrscan.io/dashboard}.} Second, the same identifier assigns every child execution to its parent without error: TWAP fills carry the \texttt{twapId} of the parent and a zero transaction hash (\texttt{0x00\ldots00}). A native TWAP therefore combines ex-ante observability with ex-post traceability.

The protocol executes the first child at the first whole second at least one second after the block that records the parent order, and the following ones every 30 seconds on whole seconds, so the first child executes between one and two seconds after the announcement (median \(1.5\) seconds). In that interval the midquote moves against the order by about \(0.1\) basis points on average (Section~\ref{subsec:ap-liquidity-provision}). We start a visible metaorder in the block that records its parent order and a hidden metaorder at its first child trade. Prices are midquotes of the order book: the start price is the last midquote before the start, and the price at any later time is the first midquote at or after that time. All price changes below are measured from the start of the metaorder, so for visible metaorders they include the move between the announcement and the first child.

\paragraph{Hidden execution: statistically reconstructed metaorders.} We reconstruct \emph{hidden metaorders} as sequences of consecutive same-sign market trades from the same address on the same market, grouped under a splitting hypothesis when inter-arrival times fall below 30 minutes \citep{bouchaud2018trades,lillo_farmer_2004_long_memory,toth_palit_lillo_farmer_2015_persistent}. These orders are executed through proprietary methods rather than a protocol mechanism, and their intended size, horizon, and rule are not disclosed ex ante; they are inferred only after the fact from address-level flow. Child trades of visible metaorders are removed before the reconstruction, so the two samples do not overlap. Appendix~\ref{app:construction} repeats the main tests with pauses of 5, 15 and 60 minutes, and with hidden metaorders rebuilt from all fills of an address, from both market and limit orders, following \citet{moro2009market}: the cost and permanent-impact discounts keep their sign and significance. We keep liquidation orders and vault addresses in the reconstruction; Appendix~\ref{app:robustness} reports the estimates without the hidden metaorders that contain a liquidation order, and Appendix~\ref{app:traders} those without vault addresses.

After sample restrictions, the final dataset contains approximately 4.3 million hidden metaorders (retaining only reconstructions with at least 10 child orders, a stricter threshold that limits false positives from coincidental clustering) and about 465{,}000 visible metaorders (with at least 5 child orders, since protocol-level identification is unambiguous even for shorter sequences). These metaorders come from about \(50{,}000\) addresses, which are \(9\%\) of the addresses that send market orders and execute \(80\%\) of the volume of market orders. Because every TWAP plans at least ten suborders, a TWAP with five to nine executed children has almost always stopped early: in the on-chain records these are \(10\%\) of TWAPs with at least five children, \(58\%\) of them cancelled by the trader and \(31\%\) stopped for insufficient margin; Appendix~\ref{app:interrupted} studies these interrupted visible metaorders. Appendix~\ref{app:robustness} repeats the main estimates with at least ten child orders in both groups and with only the visible metaorders that ran to completion. We restrict both samples to executions completing within 24 hours, consistent with the rolling 24-hour normalization window; this cap removes 28{,}904 hidden metaorders ($0.7\%$ by count and $1.5\%$ of notional).

\subsection{Descriptive Statistics}
\label{sec:metaorder-statistics}

We characterize each execution by its order size, physical duration, traded fraction, volume-time duration, participation rate, and number of child orders. For order \(i\) on market \(m_i\), notional size, physical duration, and child-order count are
\[
Q_i:=\sum_{k\in i} p_k |q_k|,
\qquad
T_i:=t_{e,i}-t_{s,i},
\qquad
N_i:=\#\{k:k\in i\},
\]
where \(q_k\) is the signed quantity of child trade \(k\), \(p_k\) its execution price, \(T_i\) is in minutes, and the sign is encoded separately by \(\epsilon_i\) so \(Q_i>0\). Let \(V_{24h,i}\) be the notional traded on market \(m_i\) over the 24 hours from the order's start, \([t_{s,i},t_{s,i}+24h)\), and \(V^{\mathrm{exec}}_i\) the notional traded there during the execution, \([t_{s,i},t_{e,i}]\). The traded fraction \(\phi_i\) factors into the participation rate \(\eta_i\) and the volume-time duration \(F_i\) of the execution:
\[
\phi_i:=\frac{Q_i}{V_{24h,i}}=\eta_i F_i,
\qquad
\eta_i:=\frac{Q_i}{V^{\mathrm{exec}}_i},
\qquad
F_i:=\frac{V^{\mathrm{exec}}_i}{V_{24h,i}}.
\] Thus \(\eta_i\) measures aggressiveness relative to contemporaneous activity rather than to daily volume.

\begin{figure}[h!]
    \centering
    \includegraphics[width=\linewidth]{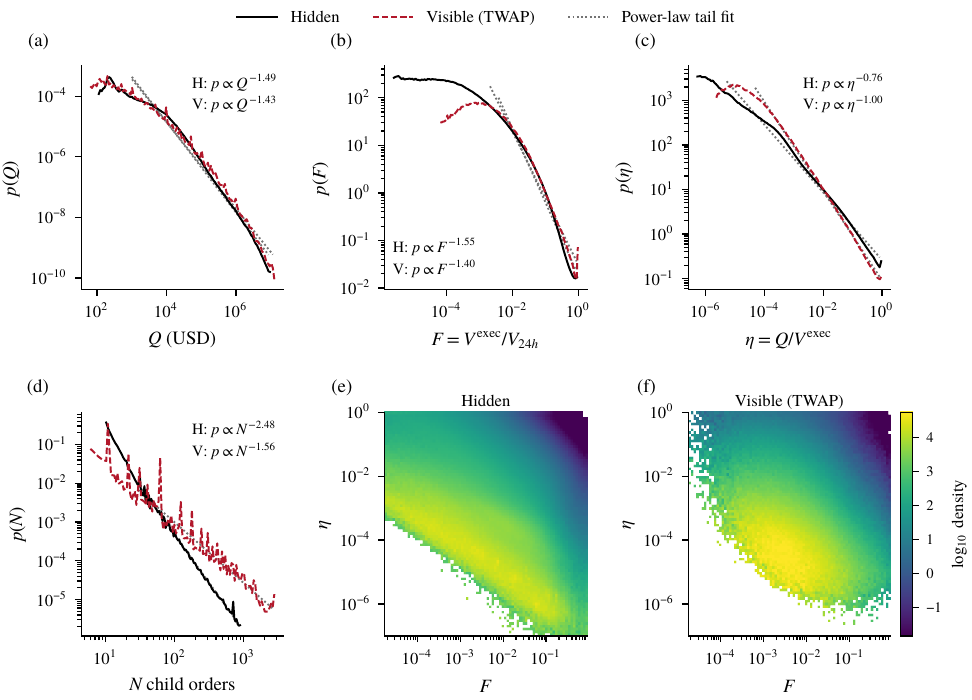}
    \caption{
    Descriptive statistics for hidden and visible metaorders.
    The top row reports the marginal distributions of order notional \(Q\),
    volume-time duration \(F=V^{\mathrm{exec}}/V_{24h}\), and participation rate
    \(\eta=Q/V^{\mathrm{exec}}\). The bottom-left panel reports the distribution of the
    number of child orders \(N\). The two remaining bottom panels report
    log-density heatmaps for the joint distribution of volume-time duration and
    participation rate, separately for hidden and visible metaorders.
    Black lines denote hidden metaorders and red lines denote visible metaorders;
    dotted grey lines are tail fits.
    }
    \label{fig:metaorder-statistics}
\end{figure}

Figure~\ref{fig:metaorder-statistics} reports the distributions of these quantities for the two samples.

The size distributions are broadly comparable in the center but differ in the right tail. The median hidden metaorder has notional about \(8.5\) thousand USD, versus \(9.1\) thousand for the median visible one; at high quantiles visible metaorders are larger, with a 95th percentile near \(496\) thousand USD against \(256\) thousand. The volume-time duration is somewhat larger for visible metaorders (median \(F\approx0.012\) versus \(0.009\); 95th percentile \(0.23\) versus \(0.16\)). The sharpest difference is in participation: median \(\eta\) is about \(3.4\%\) for hidden metaorders but only \(0.45\%\) for visible ones, and the gap persists throughout the distribution (75th percentiles of about \(20\%\) and \(5.3\%\)). Hidden metaorders are also composed of fewer child orders (median \(N=16\), 95th percentile \(73\)) than visible ones (median \(N=30\), 95th percentile \(541\)). The joint heatmaps show both populations organized around the downward-sloping envelope implied by \(\phi=\eta F\), but hidden metaorders place much more mass at high participation, while visible ones sit at lower participation and longer volume-time duration. In short, the contrast is not primarily one of order size but of \emph{execution footprint}: hidden metaorders are more concentrated and aggressive, whereas visible ones are larger in the tail, finely sliced, and executed more passively relative to contemporaneous volume.

\section{Results}
\label{sec:results}

\subsection{\texorpdfstring{Aggregate Cost and Execution Dynamics}{Aggregate Cost and Execution Dynamics}}
\label{sec:regimes}

This section compares visible and hidden metaorders in the aggregate cost of execution and in its dynamics. The two groups differ in both. These differences motivate the transparency tests and, for aggregate cost, already speak to Hypothesis~\ref{hyp:announcer}.

\subsubsection{Aggregate Execution Cost}
\label{sec:aggregate-impact}

Our measure of execution cost is temporary impact: the signed relative price change over the execution, scaled by daily volatility,
\[
\mathcal{I}_{\mathrm{tmp},i}
:=
\epsilon_i\,
\frac{S_{m_i}(t_{e,i})-S_{m_i}(t_{s,i})}
{S_{m_i}(t_{s,i})\,\sigma_{D,i}},
\qquad
\sigma_{D,i}
:=
\frac{1}{\sqrt{4\ln 2}}\,
\ln\!\left(\frac{H_i}{L_i}\right),
\]
where \(S_{m_i}(t_{s,i})\) is the start price of the metaorder (Section~\ref{sec:setting}) and \(S_{m_i}(t_{e,i})\) the midquote at completion and \(\sigma_{D,i}\) is the Parkinson daily volatility estimated over \([t_{s,i},t_{s,i}+24h)\), with \(H_i\) and \(L_i\) the high and low over that window \citep{Parkinson1980}. We estimate \(\mathcal{I}_{\mathrm{tmp}}(\phi)\) by averaging within bins of the traded fraction \(\phi_i\). This is the relation that the metaorder literature calls the square-root impact law, \(\mathcal{I}_{\mathrm{tmp}}\propto\phi^{1/2}\) \citep{toth2011anomalous,donier2015million,zarinelli2015_beyond_square_root}. Temporary impact records how far the midquote moves over the execution; it is not the average price paid on the child orders, and it leaves out fees.

\begin{figure}[h!]
  \centering
  \includegraphics[width=0.8\textwidth]{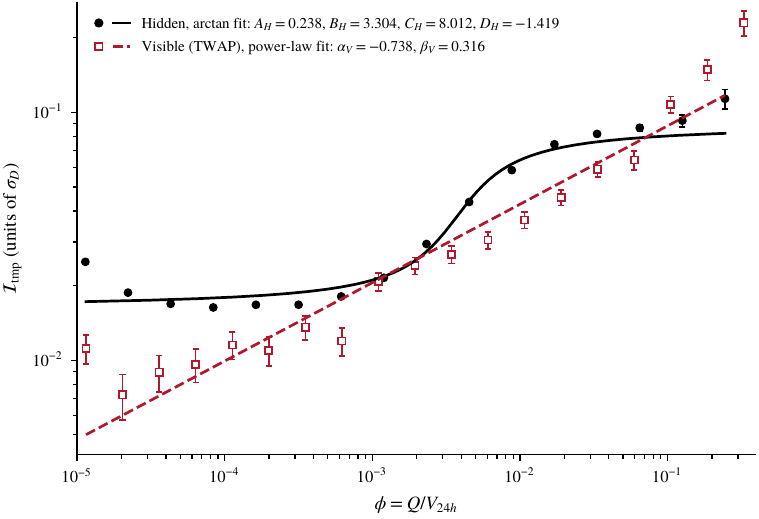}
  \caption{Temporary market impact \(\mathcal{I}_{\mathrm{tmp}}\) as a function of the
    24-hour traded fraction \(\phi = Q/V_{24h}\) for hidden and visible metaorders. Points denote binned means and error bars denote
    \(\pm 1\) standard error, on double-logarithmic axes.
    The solid black curve is the fitted relation for hidden metaorders
    \(\log_{10}(\mathcal{I}_{\mathrm{tmp}}) =
    A\,\arctan(B\log_{10}(\phi)+C)+D\); the dashed red curve is the
    fitted power law for visible metaorders
    \(\log_{10}(\mathcal{I}_{\mathrm{tmp}}) =
    \alpha+\beta\log_{10}(\phi)\).  The square-root impact law corresponds to \(\beta=1/2\). Estimates: \(\hat\alpha=-0.738\pm0.005\), \(\hat\beta=0.316\pm0.002\); \(\hat A=0.238\pm0.002\), \(\hat B=3.304\pm0.090\), \(\hat C=8.012\pm0.224\), \(\hat D=-1.419\pm0.002\).}
  \label{fig_toth0}
\end{figure}

Figure~\ref{fig_toth0} shows that visible and hidden metaorders are not described by a common cost curve. The impact of visible metaorders is close to a regular power law in \(\phi\), with a fitted exponent of about $0.32$, well below the one half of the square-root impact law. The impact of hidden metaorders is strongly curved and saturating, well captured by an arctangent form (\(R^2=0.964\)) that flattens at low \(\phi\), rises over intermediate \(\phi\), and saturates at high \(\phi\). 

Because the traded fraction \(\phi\) conflates how long an order executes with how aggressively it participates, we separate the two with \(\phi_i=\eta_i F_i\). A pooled surface \(\mathcal{I}_{\mathrm{tmp}}(\eta,F)=Y\,\eta^{\delta}F^{\gamma_1}\) has \(\hat Y=0.193\pm0.004\), \(\hat\delta=0.120\pm0.003\) and \(\hat\gamma_1=0.300\pm0.003\) (\(R^2=0.890\)): impact varies far more with volume-time duration than with participation, unlike the balanced, near-one-half exponents of \citet{zarinelli2015_beyond_square_root}. Orders with the same \(\phi\) can therefore cost differently depending on how that fraction splits between participation and duration, which is why we compare the two groups on a common support of \((\eta,F)\) in Section~\ref{sec:tests}.

\subsubsection{Execution Dynamics}
\label{sec:dynamics}

Aggregate cost compresses each order into a single number and discards the price path during execution. We therefore study impact trajectories and execution schedules. Time is normalized as \(\tau:=t/T_i\), so \(\tau=0\) is the start and \(\tau=1\) completion, and we form the signed trajectory
\[
\mathcal{I}_i(\tau)
=
\,\frac{\epsilon_i}{\sigma_{D,i}}
\left(
\frac{S_i(\tau)}{S_i(0)}-1
\right),
\qquad S_i(\tau):=S_{m_i}(t_{s,i}+\tau T_i).
\]
To compare shapes across orders of different scale, we normalize each trajectory by the fitted temporary impact predicted for its group, \(\widetilde I_i(\tau)=\mathcal{I}_i(\tau)/\widehat{\mathcal{I}}^{\,g(i)}_{\mathrm{tmp}}(\phi_i)\), using the fitted laws of Section~\ref{sec:aggregate-impact}. This removes the average size dependence within each group while leaving the completion level informative.

\begin{figure}[h!]
  \centering
  \includegraphics[width=\linewidth]{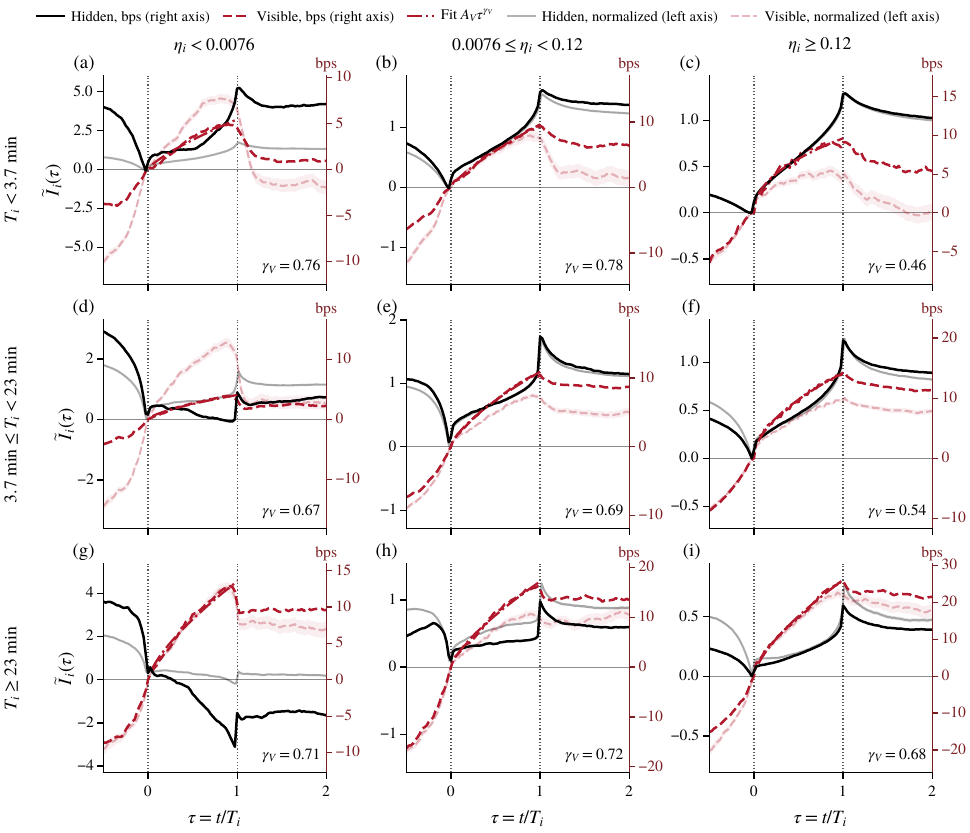}
  \caption{
  Average signed price change during execution for hidden metaorders (black solid) and visible metaorders (red dashed), by tercile of clock-time duration $T_i$ (rows) and of participation rate $\eta_i$ (columns), both increasing. The horizontal axis is $\tau=t/T_i$, from the start ($\tau=0$) to completion ($\tau=1$). Right axis: basis points; the dash-dotted red line is the power-law fit $A_V\,\tau^{\gamma_V}$ of the visible path on $\tau\in[0.05,1]$, with $\gamma_V$ reported in each panel. Left axis (faint lines): the same paths divided by the temporary impact that the fitted law of each group predicts for the order, $\widetilde I_i(\tau)=\mathcal{I}_i(\tau)/\widehat{\mathcal{I}}^{\,g(i)}_{\mathrm{tmp}}(\phi_i)$, with $\pm1$ standard error bands.  Number of metaorders (hidden / visible) per panel: (a) 255,324 / 21,152; (b) 325,399 / 8,873; (c) 624,965 / 3,988; (d) 368,378 / 95,454; (e) 408,917 / 56,988; (f) 384,513 / 37,842; (g) 486,893 / 89,382; (h) 415,227 / 33,591; (i) 176,392 / 16,128.
  }
  \label{fig:rq3_binwise_raw_perp}
\end{figure}

Figure~\ref{fig:rq3_binwise_raw_perp} conditions jointly on duration and participation. Trajectories of visible metaorders rise in a concave, strictly increasing way; in basis points they trace the smooth, square-root-like profile documented in the literature and fit well by \(A\tau^\gamma\) \citep{BacryIugaLasnierLehalle2014,zarinelli2015_beyond_square_root}. Hidden metaorders differ qualitatively: their paths have a double inflection, steep at the start and again before completion, where they peak, and flatter in between---a shape that survives when measured in basis points and is consistent with optimal execution under transient price impact \citep{ObizhaevaWang2013,gatheral2012transient}. The pre-execution segments reveal a further difference in \emph{entry}: hidden metaorders begin on average against the recent move (signed pre-start 30-minute return \(-9.0\) bps), whereas visible ones begin with the trend (\(+14.3\) bps). This is a property of the entry decision, not of the states selected: measured 24 hours before the start the gap vanishes (\(-0.2\) bps, \(p=0.68\)). We control for these entry conditions in the cost tests below.

To connect trajectories to trading behavior, we study the execution schedule \(C_i(\tau):=Q_i(\tau)/Q_i\), the cumulative fraction executed by \(\tau\), and its slope, the execution rate \(v(\tau):=dC(\tau)/d\tau\), which equals one at every \(\tau\) under uniform execution.

\begin{figure}[h!]
  \centering
  \includegraphics[width=\linewidth]{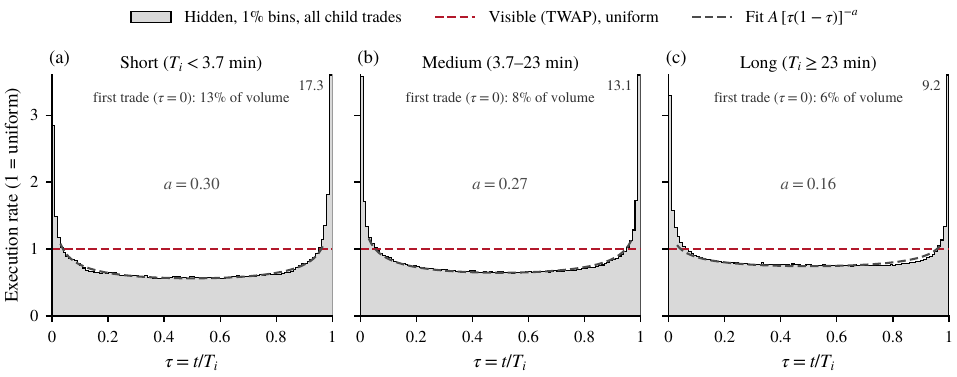}
  \caption{
  Execution rate of hidden metaorders by tercile of clock-time duration $T_i$. Each bar is the share of volume executed in $1\%$ of the order's duration, divided by $0.01$, so that $1$ is execution at a constant rate, the schedule of a visible metaorder (red line). All child trades are included: the first, at $\tau=0$, is reported in each panel and the last falls in the final bar, which is clipped (its height is written above it). The dashed grey line is the fit $v(\tau)=A[\tau(1-\tau)]^{-a}$ on $\tau\in[0.03,0.97]$, the risk-neutral optimal schedule under a power-law impact kernel (Appendix~\ref{subsec:propagator_benchmark}). 
  }
  \label{fig:execution_schedule}
\end{figure}

Figure~\ref{fig:execution_schedule} shows that the difference in trajectories is mirrored in schedules. Visible metaorders trade at a constant rate by construction. Hidden metaorders trade faster at the beginning and at the end of the order than in the middle, at every duration. An order executes \(17\%\) of its volume in the first twentieth of its duration and \(18\%\) in the last, against \(5\%\) at a constant rate. The execution rate falls from \(1.6\) in the second hundredth of the duration to \(0.66\) in the middle and rises back to \(1.6\) in the ninety-ninth. Between \(3\%\) and \(97\%\) of the duration the profile is well described by \(v(\tau)\propto[\tau(1-\tau)]^{-a}\) with \(a=0.24\) (\(R^2=0.97\)), larger for short orders (\(0.30\)) than for long ones (\(0.16\)). This is the shape of optimal execution under transient impact, discussed below. The acceleration at the end is not produced by traders who stop when the price reaches a level: it starts about a fifth of the duration before the end, and the pace near the end is the same whether the price has moved with the order or against it (Appendix~\ref{subsec:propagator_benchmark}).

Taken together, Figures~\ref{fig:rq3_binwise_raw_perp}--\ref{fig:execution_schedule} show that the near-uniform execution of visible metaorders generates smooth, concave trajectories, whereas the U-shaped schedules of hidden metaorders produce double-inflection paths that peak at completion. A transient-impact model, in which each trade moves the price by an amount that decays over time \citep{bouchaud2003fluctuations,gatheral2010no,ObizhaevaWang2013}, accounts for both shapes. A constant rate produces a concave path, and a trader who minimizes expected cost trades faster at the two ends of the horizon. The smooth U of hidden metaorders, with no flat interior, is the optimal schedule under a power-law kernel \citep{gatheral2012transient}, not under the exponential resilience of \citet{ObizhaevaWang2013}, which joins a discrete trade at each end by execution at a constant rate (Appendix~\ref{subsec:propagator_benchmark}). We do not estimate a structural model of execution. What matters for what follows is that visible and hidden metaorders differ in how they are executed, not only in whether they are announced: the cost comparisons below compare two ways of trading as a whole, not the same order with and without an announcement.

\subsection{Testing the Sunshine-Trading Hypotheses}
\label{sec:tests}

We now test the four hypotheses of Section~\ref{sec:hypotheses}.

\subsubsection{Announcer Costs (Hypothesis~\ref{hyp:announcer})}
\label{subsec:ap-announcers}

If preannouncement separates visible liquidity demand from potentially informed flow, or attracts liquidity supply when entry matters, announced orders should cost less than comparable hidden orders. We compare temporary impact on the common support of \((\eta_i,F_i)\), addressing the distributional differences of Section~\ref{sec:metaorder-statistics}. We restrict to the second and third participation columns of the \(3\times3\) sort (excluding the lowest-participation region, where signed price changes are dominated by idiosyncratic volatility), require \(F\in[3\times10^{-4},1]\), and demand at least 50 observations per group per cell of a \(25\times25\) logarithmic grid in \((\eta,F)\). The grid has \(557\) populated cells, with \(2{,}243{,}704\) hidden and \(179{,}660\) visible metaorders; in \(470\) of them both groups have positive mean temporary impact.

 Hidden metaorders have larger temporary impact than visible ones in \(83\%\) of these cells (\(81\%\) among cells with both means positive, and \(91\%\) when cells are weighted by the smaller group count), with a median impact ratio of about \(2.3\): for the same participation and duration, a hidden order moves the price about twice as much as a visible one. The gap is concentrated at short and intermediate volume-time durations: for \(F<3\times10^{-3}\) hidden metaorders exceed visible ones in every positive cell (median ratio \(\approx5.0\)); for \(3\times10^{-3}\le F<3\times10^{-2}\) the median ratio is \(\approx2.9\); and for \(F\ge3\times10^{-2}\) the two groups are close (median ratio \(\approx1.1\)). The two groups differ in how impact depends on duration and participation, not only in its level. In separable fits \(Y_g\eta^{\delta_g}F^{\gamma_g}\) to each group, the duration exponent is \(0.295\pm0.005\) for hidden and \(0.581\pm0.011\) for visible metaorders, the participation exponent \(0.149\pm0.005\) against \(0.055\pm0.011\), and the prefactor \(0.206\pm0.006\) against \(0.293\pm0.012\) (\(R^2=0.896\) and \(0.869\)): for visible metaorders, conditional on duration, participation carries little additional cost information.

To confirm the comparison is not driven by the two groups occupying different regions of the \((\eta,F)\) plane, we estimate an order-level interaction regression on the common-support sample, using median-centered log covariates so the indicator of visible metaorders is evaluated at the median execution profile. For \(x\in\{Q,F,\eta,\sigma\}\), let \(x_i^c=\log x_i-\operatorname{med}_{j\in\mathcal{S}}(\log x_j)\) and estimate
\begin{equation}
\begin{aligned}
\mathcal{I}_{\mathrm{tmp},i}
=&\ \alpha+\theta D_i^{V}
+\beta_Q Q_i^c+\beta_F F_i^c+\beta_\eta \eta_i^c+\beta_\sigma \sigma_i^c \\
&+D_i^{V}
\left(
\kappa_Q Q_i^c+\kappa_F F_i^c+\kappa_\eta \eta_i^c+\kappa_\sigma \sigma_i^c
\right)
+\mu_{m(i)}+\lambda_{h(i)}+\nu_{d(i)}+u_i ,
\end{aligned}
\label{eq:itmp-tradeoff-regression}
\end{equation}
where \(D_i^{V}\) is an indicator of visible metaorders, \(\mu,\lambda,\nu\) are market, start-hour, and start-day fixed effects, and standard errors are clustered by start day. The coefficient \(\theta\) is the gap between visible and hidden metaorders at the median profile, net of market, hour-of-day, and day effects.

The estimated gap is \(\hat{\theta}=-0.0202\) (s.e.\ \(0.0009\)), with temporary impact winsorized at the 1st and 99th percentiles as permanent impact below (\(-0.0204\) without winsorization); at the pooled median volatility \(\sigma\simeq0.049\) this is about \(10.0\) basis points lower temporary impact for a visible metaorder than for a comparable hidden one. Because the two groups start under opposite price conditions (Section~\ref{sec:dynamics}), part of the gap could in principle reflect trend continuation; adding the signed pre-start return leaves the estimate essentially unchanged at \(\hat{\theta}=-0.0199\) (\(\approx9.8\) bps). The announcer discount is not an artifact of entry conditions. The discount is not confined to a few accounts: with one weight per address it is \(7.0\) basis points, and among the addresses outside the \(1\%\) most active the same address pays \(6.4\) basis points less when it announces than when it executes hidden (\(4.1\) over all addresses, where it is less precisely estimated; Appendix~\ref{app:traders}).

\begin{table}[h!]
\centering
\begin{tabular}{lcc}
\toprule
Regressor & Hidden & Visible \\
\midrule
\(\log Q\)      & \(0.00923^{***}\) \((0.00038)\) & \(0.00789^{***}\) \((0.00041)\) \\
\(\log F\)      & \(0.00415^{***}\) \((0.00046)\) & \(0.00567^{***}\) \((0.00070)\) \\
\(\log \eta\)   & \(0.00282^{***}\) \((0.00044)\) & \(-0.00027\) \((0.00053)\) \\
\(\log \sigma\) & \(-0.01604^{***}\) \((0.00089)\) & \(-0.00035\) \((0.00115)\) \\
\midrule
Visibility indicator, \(\theta\) & \multicolumn{2}{c}{\(-0.0202^{***}\) \((0.0009)\)} \\
\bottomrule
\end{tabular}
\caption{
Order-level regression slopes for volatility-normalized temporary impact (winsorized at the 1st and 99th percentiles) on the common support defined in the text. The first column
reports \(\beta_x\), the slope for hidden metaorders; the second reports
\(\beta_x+\kappa_x\), the corresponding slope for visible metaorders. The last row reports \(\theta\), the gap between visible and hidden metaorders at the median profile. The sample contains \(2{,}243{,}704\) hidden and \(179{,}660\) visible metaorders. Standard errors,
clustered by start day, are in parentheses. Significance levels: \(^{*}p<0.05\), \(^{**}p<0.01\), \(^{***}p<0.001\).
}
\label{tab:itmp-tradeoff-regression}
\end{table}

Table~\ref{tab:itmp-tradeoff-regression} shows the two groups also differ in the margins along which cost varies. For hidden metaorders, impact rises significantly with size, duration, and participation; for visible ones it rises with size and duration but the participation slope is indistinguishable from zero. Hidden metaorders are thus penalized for executing aggressively relative to contemporaneous volume, whereas the costs of visible metaorders are driven by size and exposure over volume time. Overall, the evidence supports Hypothesis~\ref{hyp:announcer}: visible metaorders cost less than comparable hidden ones over most of the common support, especially for executions concentrated in volume time, and the two groups differ in their cost elasticities, not by a simple level shift.

\subsubsection{Information Content of Announced Flow (Hypothesis~\ref{hyp:information})}
\label{subsec:ap-information}

If announced flow is less informed, it should leave a smaller permanent price impact. The impact that persists after a trade measures its information content \citep{Hasbrouck1991}, and the impact of institutional orders placed for liquidity reasons reverts almost completely after execution, while that of other orders does not \citep{waelbroeck2013market}. We use post-execution price dynamics as a reduced-form measure of information content: temporary impact at completion mixes transitory liquidity pressure, which reverts, and information, which persists.

\begin{figure}[h!]
    \centering
    \includegraphics[width=\linewidth]{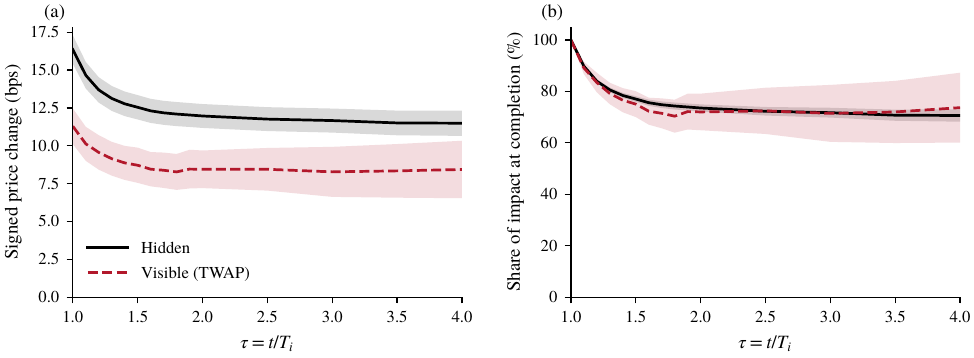}
\caption{Price path after completion for visible metaorders (red) and hidden metaorders (black). (a) Signed price change since the start of the metaorder, in basis points, measured with midquotes. (b) The same change as a share of the impact at completion of each group.  The horizontal axis is \(\tau=t/T_i\), with completion at \(\tau=1\). Averages over the six duration--participation cells of Table~\ref{tab:persistent-impact-discount}, with common weights for the two groups; \(1\%\) winsorized at each horizon; orders with prices observed up to \(\tau=4\). Bands are \(95\%\) confidence intervals from a bootstrap over start days.}
\label{fig:impact_decay}
\end{figure}

Figure~\ref{fig:impact_decay} follows the price after completion. In basis points the path of visible metaorders lies below that of hidden ones at every horizon (panel a): \(11.3\) basis points at completion and \(8.4\) at \(\tau=4\), against \(16.4\) and \(11.5\). Most of the reversion takes place within half a duration of completion. As a share of their own impact at completion, the two groups keep a similar fraction (panel b): \(74\%\) for visible and \(71\%\) for hidden metaorders at \(\tau=4\) (difference \(3\) percentage points, s.e.\ \(7\)), close to the two thirds of their peak impact that metaorders keep by the end of the day in equity markets \citep{moro2009market,bershova2013non,zarinelli2015_beyond_square_root,bucci2018slow}. The two groups thus differ in the level of the path, not in the share that reverts. This is consistent with the lower information content of announced orders being priced while they execute, rather than learned after they end. In the 50 most liquid markets, which hold three quarters of visible metaorders, visible metaorders keep less of their impact soon after completion (\(65\%\) against \(79\%\) at \(\tau=1.5\); difference \(13\) points, s.e.\ \(4\)). 

To compare orders of similar size and duration, we regress the permanent price impact on the visibility indicator. We measure the permanent price impact at post-execution horizons in multiples of the order's own duration, \(\tau\in\{1.5,2,3,4\}\),
\begin{equation}
    I_i^{\mathrm{perm}}(\tau)
    =
    10^4\,\epsilon_i
    \frac{S_i(\tau)-S_i(0)}{S_i(0)} ,
    \label{eq:persistent-impact}
\end{equation}
and estimate, for each horizon,
\begin{equation}
    I_i^{\mathrm{perm}}(\tau)
    =
    \alpha
    + \beta_{V}D_i^{V}
    + c_1\sqrt{\phi_i}
    + c_2\eta_i
    + c_3\log T_i
    + c_4\sigma_{D,i}
    + \mu_{m(i)}+\lambda_{h(i)}+\nu_{d(i)}
    + u_i ,
    \label{eq:persistent-impact-regression}
\end{equation}
and the same specification without the traded-fraction control \(\sqrt{\phi_i}\). Fixed effects are for market, start hour, and start day, as in Eq.~\eqref{eq:itmp-tradeoff-regression}; standard errors are clustered by day. The sample is restricted to participation rates of at least \(0.0076\), the lower end of the common support of Table~\ref{tab:itmp-tradeoff-regression}. The dependent variable is winsorized at the 1st and 99th percentiles. A negative \(\beta_{V}\) means visible metaorders leave a smaller permanent price impact.

\begin{table}[htbp]
\centering
\begin{tabular}{lcccc}
\toprule
Spec. & \(\tau=1.5\) & \(\tau=2\) & \(\tau=3\) & \(\tau=4\) \\
\midrule
Eq.~\eqref{eq:persistent-impact-regression}
  & \(-4.80^{***}\) & \(-4.36^{***}\) & \(-4.54^{***}\) & \(-4.40^{***}\) \\
  & \((0.73)\)      & \((0.73)\)      & \((0.87)\)      & \((0.97)\)      \\[4pt]
Without \(\sqrt{\phi}\)
  & \(-2.72^{***}\) & \(-2.33^{***}\) & \(-2.58^{**}\) & \(-2.53^{**}\) \\
  & \((0.67)\)      & \((0.68)\)      & \((0.84)\)      & \((0.95)\)      \\
\bottomrule
\end{tabular}
\caption{Permanent-impact discount of visible metaorders. The table reports
the coefficient \(\beta_{V}\) on \(D_i^{V}\) from
Eq.~\eqref{eq:persistent-impact-regression} and from the same specification without \(\sqrt{\phi}\).
The dependent variable is \(I_i^{\mathrm{perm}}(\tau)\), measured in basis
points. Standard errors, clustered by calendar day, are in parentheses. All
specifications include market, start-hour, and start-day fixed effects. The
sample contains approximately \(2.66\) million observations and \(225\) day
clusters. \({}^{**}p<0.01\), \({}^{***}p<0.001\).}
\label{tab:persistent-impact-discount}
\end{table}

Table~\ref{tab:persistent-impact-discount} shows visible metaorders leave a smaller permanent price impact at every horizon: between \(-4.4\) and \(-4.8\) basis points in the baseline, and \(-2.3\) to \(-2.7\) without the traded-fraction control. With errors clustered by address as well as by day, the discount is borderline significant (\(t=-1.9\) at \(\tau=1.5\) and \(-1.7\) at \(\tau=4\); Appendix~\ref{app:traders}). For scale, the average impact at completion is \(16.4\) basis points for hidden metaorders and \(12.2\) for visible ones, and the average permanent price impact at \(\tau=4\) is \(11.5\) and \(9.3\) basis points: the discount is about two fifths of the permanent price impact of a hidden order. Table~\ref{tab:levels-cells} reports both levels by duration and participation. Since hidden metaorders are on average contrarian, part of their larger permanent footprint could reflect reversal of the pre-entry dip; controlling for the signed pre-start return attenuates the coefficient only by \(5\)--\(11\%\) (e.g.\ from \(-4.8\) to \(-4.6\) at \(\tau=1.5\)), leaving the conclusion intact.

The price of the market keeps moving after completion, and this movement adds noise to both comparisons. We remove it by comparing visible and hidden metaorders that trade the same market in the same direction and end within the same hour, so that after completion they face the same market movement. Within these groups, and with the same controls for duration and participation, at completion the price has moved about \(8\) basis points less for visible metaorders than for hidden ones (s.e.\ \(0.7\)), and the difference stays between \(8\) and \(10\) basis points up to five days after completion (s.e.\ \(0.6\) to \(1.3\)). It is similar in the two halves of the sample and when each of the ten markets with most groups is left out in turn.

The cost discount of Table~\ref{tab:itmp-tradeoff-regression} and the permanent-impact discount of Table~\ref{tab:persistent-impact-discount} keep their sign and significance with transaction prices, with volume and volatility measured over the 24 hours before the start, without the hidden metaorders near liquidations, with at least ten child orders in both groups, and with only the visible metaorders that ran to completion (Appendix~\ref{app:robustness}).

Consistent with Hypothesis~\ref{hyp:information}, the price impact that survives execution, the component most plausibly tied to information, is smaller for visible metaorders.

\subsubsection{Nonannouncer Costs (Hypothesis~\ref{hyp:nonannouncer})}
\label{subsec:ap-nonannouncers}

We now ask whether hidden traders bear a higher cost because part of the flow is announced, the mirror of the announcer benefit. In the model of \citet{AdmatiPfleiderer1991}, announcing identifies a liquidity trader as uninformed. The flow that stays unannounced is then more likely to be informed, the price responds to it more, and liquidity traders who do not announce pay more. This cost depends on how much liquidity demand a market announces, and it is small when few announce: it is at most the announced share of liquidity demand times the difference between the price responses to unannounced and to announced flow. Visible metaorders execute \(3.8\%\) of taker volume and the cost gap is about ten basis points (Section~\ref{subsec:ap-announcers}), which gives an illustrative value of \(0.4\) basis points, and \(1.4\) if three quarters of unannounced volume is informed (Appendix~\ref{app:bound}).

In the model the cost shifted onto nonannouncers grows with the share of flow that is announced. We therefore test whether hidden metaorders cost more in the months in which a larger share of their market's flow is announced. For hidden metaorder \(i\) we estimate
\begin{equation}
    I_i^{\mathrm{perm}}(\tau) = \beta\,a_{m(i),p(i)} + \mathbf{c}^{\prime}\mathbf{x}_i + \mu_{m(i)}+\lambda_{h(i)}+\nu_{d(i)} + u_i ,
    \label{eq:h3-share}
\end{equation}
where \(I_i^{\mathrm{perm}}(\tau)\) is the price change of Eq.~\eqref{eq:persistent-impact}, at completion (\(\tau=1\)) and at \(\tau=4\), \(a_{m,p}\) is the share of taker volume executed through visible metaorders in market \(m\) and calendar month \(p\), \(\mathbf{x}_i=(\log Q_i,\eta_i,\log T_i,N_i)\) are the size, participation rate, duration and number of child orders of the metaorder, and the fixed effects are those of Eq.~\eqref{eq:persistent-impact-regression}; standard errors are clustered by market. With market fixed effects, the coefficient compares the same market in months in which more or less of its flow is announced. The sample contains about \(2.3\) million hidden metaorders in \(199\) markets. An increase of ten percentage points in the announced share corresponds to \(+0.2\) basis points of impact at completion (s.e.\ \(0.9\)) and \(+0.1\) of permanent price impact at \(\tau=4\) (s.e.\ \(1.1\)). Neither estimate is distinguishable from zero: we detect no cost shifted onto hidden metaorders, though we cannot exclude a small one.

The previous test compares different months of the same market. A second test looks at the moment of trading: does a hidden metaorder cost more when visible metaorders are executing in its market when it starts? It does when they trade in its direction, but this is the price pressure of that flow, not a cost of its announcement. We measure the flow running when a hidden metaorder starts by the execution rates of the other metaorders active at that moment, split into visible and hidden, same and opposite direction, and estimate
\begin{equation}
    I_i^{\mathrm{perm}}(\tau) = \beta_{V+}\,s_i^{V+} + \beta_{V-}\,s_i^{V-} + \beta_{H+}\,s_i^{H+} + \mathbf{c}^{\prime}\mathbf{x}_i + \mu_{m(i)}+\lambda_{h(i)}+\nu_{d(i)} + u_i ,
    \label{eq:h3-flows}
\end{equation}
where \(s_i^{g}\) is the share of group \(g\) in the total execution rate of the other metaorders active when \(i\) starts, \(V\) and \(H\) denote visible and hidden metaorders, \(+\) and \(-\) those in the same and in the opposite direction of \(i\), and opposite-direction hidden flow is the omitted group; controls and fixed effects are those of Eq.~\eqref{eq:h3-share}, and standard errors are clustered by day. Same-direction visible flow raises the cost at completion by \(\hat\beta_{V+}=8.7\) basis points per unit share (s.e.\ \(0.6\)), and same-direction hidden flow by \(\hat\beta_{H+}=8.0\) (s.e.\ \(0.5\)). The difference, \(0.8\) (s.e.\ \(0.7\)), is not distinguishable from zero, and neither is the difference on the opposite side, \(\hat\beta_{V-}=0.9\) (s.e.\ \(0.7\)): we do not detect an effect of announced flow on the cost of hidden metaorders trading alongside it. The on-chain TWAP records let us measure visible flow by the size announced and not yet executed, which the market knows from the announcement; the difference is then \(0.9\) (s.e.\ \(1.0\)). Without the \(1\%\) most active addresses, visible flow adds more price pressure than hidden flow on both sides, and the part of its effect that does not depend on the side is not positive (Appendix~\ref{app:traders}).

We thus detect no robust effect of announcement on the costs of nonannouncers: the estimates do not determine its sign.

\subsubsection{Liquidity Provision around Announced Orders (Hypothesis~\ref{hyp:liquidity})}
\label{subsec:ap-liquidity-provision}

The final prediction concerns liquidity supply. Two channels point the same way: with costly entry, suppliers can condition participation on realized announced demand; even without an entry margin, suppliers who infer that announced flow is relatively uninformed face less adverse selection on that side and can quote more aggressively. Since \citet{AdmatiPfleiderer1991} model liquidity supply in a stylized batch auction, we adapt the prediction to a CLOB, where we directly observe displayed depth, imbalance, and resting orders.

We retain events of at least five minutes and \(10{,}000\) USD notional (and, for hidden metaorders, at least ten child trades), and study relative spreads, displayed depth, order-book imbalance, and fixed-notional sweep costs. All side-dependent variables are oriented by order direction, so positive imbalance, higher depth, and lower sweep cost all mean more liquidity available to absorb the flow. For a snapshot, letting \(B^{(k)}\) and \(A^{(k)}\) be displayed bid and ask size within \(k\) quoted spreads of the mid, oriented imbalance is
\[
    \mathrm{Imb}^{(k)}
    =
    s\,\frac{B^{(k)} - A^{(k)}}{B^{(k)} + A^{(k)}} ,
    \qquad
    s =
    \begin{cases}
    -1, & \text{buy order},\\
    \phantom{-}1, & \text{sell order},
    \end{cases}
\]
with \(k=5\); depth is the corresponding dollar depth on the absorbing side, relative spread is \((a^{1}-b^{1})/\mathrm{mid}\) in basis points, and sweep cost is the extra cost of immediately executing a fixed notional on the absorbing side. Under the sunshine-trading reading, a liquidity-supply response should appear away from the touch---imbalance up, depth up, sweep cost down---while the quoted spread is ambiguous, since suppliers may widen the inside quote to protect against a predictable execution even as they add depth behind it.

\begin{figure}[h!]
    \centering
    \includegraphics[width=\linewidth]{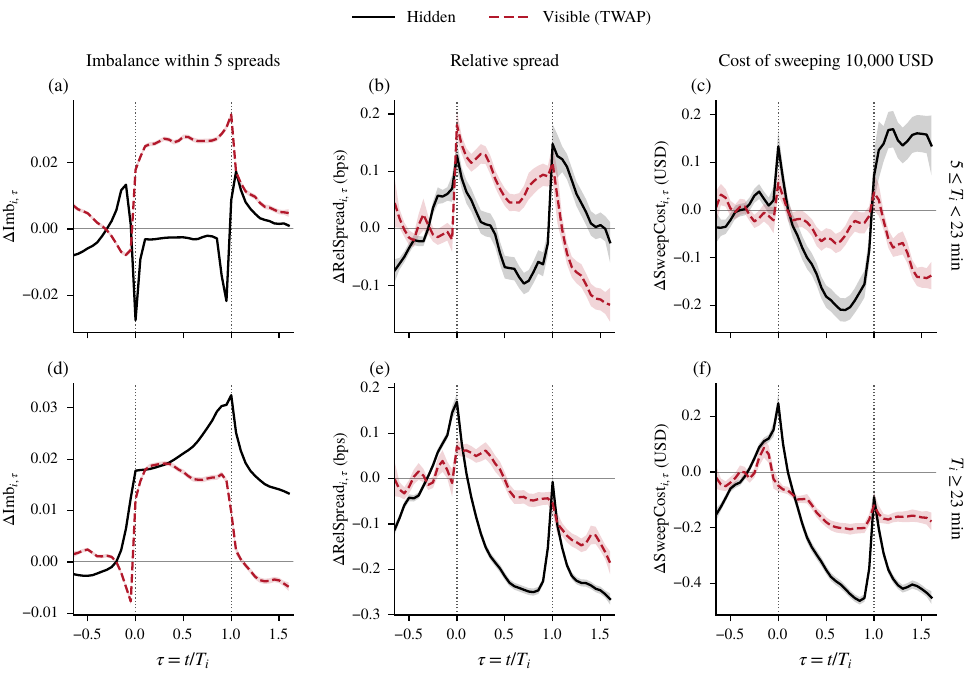}
    \caption{
    Order book around visible metaorders (red dashed) and hidden metaorders (black solid) lasting 5 to 23 minutes (top) and more than 23 minutes (bottom). Columns: oriented imbalance within five quoted spreads of the midquote; relative spread, in basis points; cost of sweeping \(10{,}000\) USD on the side of the book the order takes from, in USD, that is, in basis points of the amount traded. \(\tau=t/T_i\), with \(\tau=0\) the start (for visible metaorders, the parent order) and \(\tau=1\) completion. Changes from the event's own pre-window mean, in bins of \(0.05\) order durations, smoothed with a three-bin moving average within each phase; bands are \(\pm 1\) standard error.
    }
    \label{fig:twap-liquidity-response}
\end{figure}

Figure~\ref{fig:twap-liquidity-response} compares book dynamics around visible and hidden metaorders. For orders of 5 to 23 minutes the oriented imbalance jumps toward the absorbing side when a visible metaorder is announced and stays there until completion: the book tilts toward a visible metaorder as soon as it is public. Around hidden metaorders of the same duration it does not rise. For longer orders the book tilts toward hidden metaorders as well, and the tilt grows during execution, probably because the market learns about the order from the repeated trades of the same address. Around visible metaorders the inside quote is slightly wider than around hidden ones, while the cost of sweeping \(10{,}000\) USD falls for both.

Because the figure is descriptive, we complement it with an event-time placebo regression. For event \(i\) and relative minute \(m\),
\begin{equation}
    y_{i,m}
    =
    \alpha_i
    + \lambda_m
    + \beta\,D_i^{V}\,\mathbf{1}\{0 \leq m \leq T_i\}
    {}+ \gamma\,D_i^{V}\,\mathbf{1}\{m > T_i\}
    + u_{i,m},
    \label{eq:liquidity-response}
\end{equation}
where \(y_{i,m}\) is a book measure averaged over the snapshots of relative minute \(m\), \(\alpha_i\) are event fixed effects (one for each visible metaorder and one for each placebo), \(\lambda_m\) relative-minute fixed effects, and the treatment equals one only for visible metaorders during their active window; ten placebo starts for each visible metaorder, drawn at random times in the same market and day, form the comparison group, and standard errors are clustered by visible metaorder together with its placebos. The panel runs from 30 minutes before to 60 minutes after the start; with the post-completion indicator, \(\beta\) is the change during execution relative to the event's own pre-window, net of the change at placebo events in the same relative minute. In two further specifications we add to Eq.~\eqref{eq:liquidity-response} a pre-window lead \(\delta\,L_{i,m}\), with \(L_{i,m}=D_i^{V}\,\mathbf{1}\{-10\le m<0\}\), which compares the last ten minutes before the start with the preceding twenty, to check for differential pre-trends, and an interaction \(\zeta\,A_{i,m}\,q_i\), where \(A_{i,m}=D_i^{V}\,\mathbf{1}\{0\le m\le T_i\}\) is the active-window indicator and \(q_i=\log(Q_i/Q_{\mathrm{med}})\) is log executed notional relative to its median (about \(50{,}000\) USD), to test whether the response scales with executed size.

\begin{table}[h!]
\centering
\begin{tabular}{l r@{\;}l r@{\;}l r@{\;}l}
\toprule
Book outcome & \multicolumn{2}{c}{Active window $\hat\beta$} & \multicolumn{2}{c}{Pre-window lead $\hat\delta$} & \multicolumn{2}{c}{Size interaction $\hat\zeta$} \\
\midrule
Relative spread (bps) & $0.050^{**}\phantom{^{*}}$ & $(0.016)$ & $0.021\phantom{^{***}}$ & $(0.022)$ & $-0.013^{*}\phantom{^{**}}$ & $(0.006)$ \\
Imbalance, $5\times$ spread & $0.0170^{***}$ & $(0.0004)$ & $-0.0063^{***}$ & $(0.0004)$ & $0.0013^{***}$ & $(0.0002)$ \\
Depth, $5\times$ spread (USD) & $4221^{***}$ & $(737)$ & $-5899^{***}$ & $(779)$ & $1850^{***}$ & $(527)$ \\
Sweep cost, $10{,}000$ USD (bps) & $-0.0648^{***}$ & $(0.0153)$ & $0.0841^{**}\phantom{^{*}}$ & $(0.0308)$ & $0.0316^{***}$ & $(0.0056)$ \\
\bottomrule
\end{tabular}
\caption{Order book during the execution of visible metaorders, all durations. \(\hat\beta\) is the active-window effect from Eq.~\eqref{eq:liquidity-response}, \(\hat\delta\) the pre-window lead (last ten minutes before the start relative to the preceding twenty) and \(\hat\zeta\) the size interaction. Imbalance and depth are measured over the book levels within five quoted spreads of the midquote, up to the tenth level on each side. About \(181.6\) million event-minute observations, from \(183{,}743\) visible metaorders and \(1{,}837{,}370\) placebo events. Standard errors, clustered by visible metaorder together with its placebos, in parentheses. Significance: \(^{*}p<0.05\), \(^{**}p<0.01\),
\(^{***}p<0.001\).}
\label{tab:twap-liquidity-response-regression}
\end{table}

The book responds to an announcement by tilting toward the announced order, not with a tighter quote (Table~\ref{tab:twap-liquidity-response-regression}). During the active window, relative to the event's own pre-window and net of placebo events, the oriented imbalance rises by \(0.017\) and displayed depth on the absorbing side is about \(4{,}200\) USD higher, while the inside spread is slightly wider. Both responses grow with the executed size: larger visible metaorders thus meet a book that tilts further toward them, as the model predicts when liquidity suppliers enter after seeing the announced size. The book is not already tilting before the start: in the last ten minutes imbalance and depth are lower and the sweep cost is higher, while the spread does not change detectably, so visible metaorders tend to start after a short deterioration of liquidity, in line with their trend-following entry. During execution the cost of sweeping \(10{,}000\) USD is lower than in the pre-window, and depth and imbalance remain higher when the comparison is with its earlier, calmer minutes. After completion the imbalance falls below its pre-window level (\(\hat\gamma=-0.0083\), s.e.\ \(0.0005\)): the tilt lasts only as long as the program. This supports Hypothesis~\ref{hyp:liquidity}: the book tilts toward visible execution, and the response scales with the executed size.

\paragraph{Increased liquidity supply or decreased liquidity demand?} A book that tilts toward the absorbing side need not mean that suppliers add liquidity there: the same pattern would arise if other traders consumed less of it, for instance because same-direction traders step back once a program is announced. We therefore estimate Eq.~\eqref{eq:liquidity-response} on the taker flow of all traders other than visible metaorders, measured in the 60 seconds before each minute of the panel and oriented by the direction of the event (Table~\ref{tab:other-taker-flow}).

\begin{table}[h!]
\centering
\begin{tabular}{l r@{\;}l r@{\;}l r@{\;}l}
\toprule
Outcome & \multicolumn{2}{c}{Active window $\hat\beta$} & \multicolumn{2}{c}{Pre-window lead $\hat\delta$} & \multicolumn{2}{c}{Size interaction $\hat\zeta$} \\
\midrule
Taker volume, same (log) & $-0.058^{***}$ & $(0.003)$ & $0.174^{***}$ & $(0.003)$ & $0.047^{***}$ & $(0.002)$ \\
Taker volume, opposite (log) & $0.075^{***}$ & $(0.003)$ & $0.102^{***}$ & $(0.003)$ & $0.062^{***}$ & $(0.002)$ \\
Share of taker volume, same & $-0.0139^{***}$ & $(0.0003)$ & $0.0101^{***}$ & $(0.0004)$ & $-0.0017^{***}$ & $(0.0002)$ \\
Hidden starts, same & $-0.070^{***}$ & $(0.003)$ & $0.045^{***}$ & $(0.004)$ & $-0.000\phantom{^{***}}$ & $(0.002)$ \\
Hidden starts, opposite & $-0.023^{***}$ & $(0.003)$ & $0.055^{***}$ & $(0.004)$ & $0.015^{***}$ & $(0.002)$ \\
\bottomrule
\end{tabular}
\caption{Taker flow of other traders around visible metaorders. Same
specification, sample, and reporting conventions as
Table~\ref{tab:twap-liquidity-response-regression}. Taker volume is the USD
volume of market orders that are not children of any visible metaorder, in the 60 seconds before each minute of the panel, entered as \(\log(1+\mathrm{USD})\);
``same'' is the direction of the visible metaorder; the share is defined in the minutes in which other traders have taker volume (\(174.0\) million observations). Hidden starts count hidden metaorders (Section~\ref{sec:setting}) that begin in the same
window; their pre-window means around visible metaorders are both \(0.62\) per minute. Significance: \(^{*}p<0.05\), \(^{**}p<0.01\), \(^{***}p<0.001\).}
\label{tab:other-taker-flow}
\end{table}

Table~\ref{tab:other-taker-flow} shows three things. First, visible metaorders start when the market is busy: in the ten minutes before the start, other traders' taker volume is \(11\)--\(19\%\) higher than in the preceding twenty, on both sides. Second, once the program is public, other traders do not trade with it: the share of their taker volume in its direction falls by about \(1.4\) percentage points, from \(50.9\%\), and \(9\)--\(11\%\) fewer hidden metaorders start in its direction. This is the opposite of what anticipatory or predatory trading would produce. Third, the fall in that share is larger for larger programs, while the drop in hidden starts does not depend on size. The clearest result on other traders is thus that their share in the direction of the visible metaorder falls. This is compatible with lower liquidity consumption on the absorbing side, which by itself would tilt the book toward the program, but it does not show that suppliers add orders there.

These results are stable to how the placebo events are drawn and to how imbalance is measured. With placebo starts matched to each visible metaorder on the hour of day and on the spread and volatility of the preceding 30 minutes, the changes in imbalance, depth, sweep cost and the same-direction share of other traders' taker volume are consistent with those of Tables~\ref{tab:twap-liquidity-response-regression} and~\ref{tab:other-taker-flow}. The rise in imbalance also holds within fixed bands of 10 and 25 basis points around the midquote, and for visible metaorders shorter and longer than 23 minutes.

\paragraph{Is there evidence of front-running?} A visible metaorder is public one to two seconds before its first child executes, and its later children arrive every 30 seconds, which leaves room to front-run it, that is, to trade in its direction before a child and unwind afterwards. \citet[Section~4.6]{AdmatiPfleiderer1991} do not model front-running, but argue that it should not hurt a trader whose announcement is public: the trades of front-runners are anticipated from the announcement and should not move the price, and a front-runner who finds a counterparty supplies liquidity to it. Our results agree with this argument. On average, before the first child the market orders of other traders in the direction of the order and the depth on the side of the book that the child takes do not differ from those at random times of the same market and day, and the midquote moves by about \(0.06\) basis points net of the trend of the preceding ten seconds. Before later children none of the three differs from the same order 15 seconds earlier. Over the whole execution other traders trade less, not more, in the direction of the order (Table~\ref{tab:other-taker-flow}), and visible metaorders cost less than comparable hidden ones net of any front-running (Section~\ref{subsec:ap-announcers}).

\section{Robustness and Additional Analysis}
\label{sec:robustness}

Several concerns qualify the interpretation of these results. Some can be addressed with the data, whereas others limit what the comparison can show.

\paragraph{Self-selection into visible execution.} The contrast between visible and hidden metaorders is observational: traders choose whether to announce, and the choice depends mostly on who the trader is. Out of sample, the share of visible metaorders in a market predicts the choice poorly (AUC \(0.674\)) and the state of the book before the start adds little (\(0.685\)), whereas the share of visible metaorders among the other orders of the same address predicts it almost perfectly (\(0.998\)): \(86.6\%\) of orders come from addresses that always execute in the same way. An address is not necessarily a trader. On Hyperliquid a trader can open sub-accounts, each with its own address, and the protocol reports the master account they belong to; grouping sub-accounts under their master accounts into entities lowers this share only to \(85.4\%\). A trader could also use addresses that are not linked to each other, which we cannot group, but this is costly: trading fees fall with the volume of an account over the previous 14 days.

The separation is asymmetric. Of the \(48{,}398\) entities with at least one metaorder, \(34{,}774\) execute only hidden metaorders, \(6{,}450\) only visible ones, and \(7{,}174\) both. Those who execute hidden metaorders almost never announce: \(93\%\) of hidden metaorders come from entities that never executed a visible one. The converse does not hold: only \(17\%\) of visible metaorders come from entities that use nothing else, and about half of visible volume comes from entities with between \(5\%\) and \(95\%\) of their orders in visible form. This allows a comparison within trader. With entity fixed effects in Eq.~\eqref{eq:persistent-impact-regression}, the permanent-impact discount of visible metaorders is \(-6.7\) basis points (s.e.\ \(1.4\)) at \(\tau=1.5\) and \(-7.0\) (s.e.\ \(2.4\)) at \(\tau=4\), against \(-4.8\) and \(-4.4\) without them (Appendix~\ref{app:traders}): the same trader leaves a smaller permanent impact when announcing, so the result of Section~\ref{subsec:ap-information} is not only a difference between types of traders. Within these entities the choice depends on size rather than on market conditions: doubling the notional lowers the probability of announcing by about six percentage points, while the return before the start and volatility matter little. Nor do visible metaorders cluster in time: they are about \(10\%\) of all orders on a typical day, and only slightly fewer on volatile days.

Our estimates therefore compare visible and hidden execution as traders choose them, not the same order executed in two ways. This is the relevant comparison for a venue that offers both, but it is not a causal effect of visibility; for this reason we rely on the within-trader estimate, on the response of the book growing with size (Section~\ref{subsec:ap-liquidity-provision}), and on controls for the conditions at the start.

\paragraph{Hidden flow is a mixture.} In \citet{AdmatiPfleiderer1991} the announcer is compared with a liquidity trader who does not announce. We do not observe that trader separately: hidden metaorders pool uninformed liquidity demand with informed order splitting (Section~\ref{sec:hypotheses}), and nothing in the data identifies which is which. Two consequences follow. The \(10\) basis points of Section~\ref{subsec:ap-announcers} measure the cost of announced flow relative to this mixture, so part of the gap may reflect the information content of hidden orders rather than a saving from announcement; the model's prediction concerns the second component only. And the smaller permanent impact of visible metaorders shows that announced flow is on average less informed than hidden flow, not that hidden flow is informed. The within-trader estimate above goes some way toward the model's comparison, since it holds the trader fixed, but a trader may still choose to announce precisely the orders that are not information-motivated.

\paragraph{Why would an informed trader not announce?} If announced orders cost less, a trader with private information has a reason to announce too, and the discount should not survive. \citet[Section~4.1]{AdmatiPfleiderer1991} give two reasons why it can. The first is that announcing delays execution, which costs more to a trader whose information may become public in the meantime. The second is that announcers can be identified: traders may identify themselves when they preannounce, or the announcement may come from intermediaries who acquire a reputation for screening out informed traders. On Hyperliquid the address that sends each order is public, and so are its earlier orders, so the market could price an announced order on the record of its trader.
The pattern resembles a dealer market, in which traders who can convince a dealer that they are uninformed trade at better terms \citep{Roell1990}.
The analysis of the dynamics of the information content of TWAPs progressively sent by the same trader is left for a future study.

\paragraph{Hidden does not mean anonymous.} Hidden metaorders are not hidden in the sense of anonymous institutional flow: every trade reports an address, so same-address, same-direction trading can be inferred while it unfolds. The distinction we exploit is between explicit protocol-level preannouncement and gradual statistical inferability. This matters for the book-level test: the gradual imbalance buildup around hidden metaorders (Section~\ref{subsec:ap-liquidity-provision}) is consistent with participants learning from repeated trades, whereas visible metaorders reveal the program from inception. The hidden benchmark is therefore partially learnable, which if anything biases against finding a visibility effect.

\paragraph{Predictability of order arrivals.} The book does not announce a visible metaorder before the trader does. In monthly out-of-sample tests, the start of a visible metaorder in a market and five-minute interval is predicted by the recent activity of that market; once activity and recent taker flow are included, removing spread, depth and imbalance leaves the prediction unchanged (Appendix~\ref{app:arrivals}). The evidence of Section~\ref{subsec:ap-liquidity-provision} is about how the book responds to an announced program, not about the book predicting one.

\paragraph{Passive volume and classification.} The children of a visible metaorder are market orders, so hidden metaorders built from market orders compare executions of the same kind. An address may, however, also trade with limit orders while its hidden metaorder executes. Appendix~\ref{app:fmo} shows that this passive volume is concentrated in a few addresses that provide liquidity on both sides, and repeats the tests with hidden metaorders rebuilt from all fills of an address, from both market and limit orders, keeping those executed mostly with market orders, as in \citet{moro2009market}: both discounts keep their sign and significance.

\paragraph{A secondary venue.} For most assets in our sample, and for the largest ones in particular, prices are formed mainly on centralized exchanges, and Hyperliquid is a secondary venue. Over the same period, the USD 1.93 trillion traded in our sample compares with USD 17.0 trillion of perpetual-futures volume on Binance and USD 30.1 trillion on Binance, Bybit and OKX combined, about \(11\%\) and \(6\%\) of theirs; its average open interest, USD 4.4 billion counting one side of each position, was about \(9.5\%\) of that on the three exchanges.\footnote{Daily volume and open interest of all perpetual contracts, including delisted ones, from the exchanges' public data: Binance Public Data (\url{https://data.binance.vision}), the Bybit API v5 (\url{https://bybit-exchange.github.io/docs/v5/intro}) and the OKX API v5 (\url{https://www.okx.com/docs-v5/en/}); open interest of Hyperliquid from 0xArchive (\url{https://www.0xarchive.io}). Accessed November 19, 2026.} Two consequences follow. The price impact that persists after a metaorder reflects not only what Hyperliquid's liquidity suppliers infer from it, but also how its price is realigned with other venues, so the permanent impact we measure is a local quantity. And liquidity suppliers on Hyperliquid can hedge or take their prices from other markets, so local depth and adverse selection differ from those of the isolated market in \citet{AdmatiPfleiderer1991}. Visible and hidden metaorders trade in the same markets and are compared within market, hour and day, so these features affect both groups; we expect them to bear more on the level of costs than on the gap between the two.

\paragraph{External validity.} Hyperliquid differs from equity and futures markets in ways that may matter for our results: every order is tied to a public address, the contracts are crypto perpetual futures traded around the clock, and the same contracts trade in larger volume on centralized exchanges. It is therefore not obvious that our conclusions extend to those markets.

\section{Conclusion}
\label{sec:conclusion}
\label{sec:discussion}

On Hyperliquid a trader can execute a large order as a native TWAP, whose terms are public from inception, or split it into many trades without announcing it. Comparing the two, we test the predictions of \citet{AdmatiPfleiderer1991} on preannounced trading. Visible metaorders move the price less during execution and leave a smaller permanent impact than comparable hidden metaorders. We detect no cost shifted onto traders who do not announce. The order book tilts toward the side that absorbs a visible metaorder, more so for larger orders, and other traders trade less in its direction; we find little to no evidence of front-running. For the orders that traders choose to announce, preannouncement thus attracts liquidity more than predation.

These results matter for those who design markets, for those who trade in them, and for those who regulate them. A venue that offers an order type whose terms are public gives its users a cheaper way to execute, and liquidity suppliers respond to it. A trader who is willing to reveal an order can lower its cost: the discount is about 10 basis points during execution and 5 after it, and it holds when the same trader is compared with itself. For the debate on how transparent markets should be, the evidence is that letting traders reveal their intentions improves their execution with no cost that we can detect for those who do not.

Two limits bound these conclusions. Announcing is a choice on Hyperliquid, so we observe the orders that traders decide to reveal: a rule that required disclosure would change which orders remain hidden, what an announcement conveys, and the incentives to supply liquidity, and our evidence does not say what would follow. And the evidence comes from one market, whose traders, contracts and rules differ from those of equity and futures markets. Most traders in our sample never announce, although those who do pay less; why they do not is a question that our data cannot answer. What this market offers is a public record of both the orders that traders announce and those they do not: a rare empirical laboratory, where the costs and benefits of pre-trade transparency can be measured rather than inferred.

\ifanon\else
\section*{\texorpdfstring{Acknowledgments}{Acknowledgments}}
In preparing this paper, the authors used generative AI tools (Claude, Anthropic; GPT, OpenAI) as an aid in writing and revising the manuscript and for programming support. The authors reviewed and verified all content and take full responsibility for it.
\fi

\bibliographystyle{apalike}
\bibliography{references_MS}

\newpage
\appendix
\setcounter{table}{0}\setcounter{figure}{0}
\renewcommand{\thetable}{A\arabic{table}}\renewcommand{\thefigure}{A\arabic{figure}}
\renewcommand{\theHtable}{A\arabic{table}}\renewcommand{\theHfigure}{A\arabic{figure}}
\section{Appendix: Additional Results}
\label{app:additional}
\subsection{\texorpdfstring{Execution Schedules and the Propagator Benchmark}{Execution Schedules and the Propagator Benchmark}}
\label{subsec:propagator_benchmark}

To interpret the execution-style evidence of Section~\ref{sec:dynamics}, we use a simple transient-impact benchmark based on a propagator model \citep{bouchaud2003fluctuations,gatheral2010no}. In model time \(v\in[0,F]\),
\[
s(v)=s(0)+\int_0^v q(u)\,G(v-u)\,du + W_v,
\]
where \(q(u)\) is the signed execution rate, \(G(\cdot)\) is the transient-impact kernel, and \(W_v\) captures exogenous fluctuations. The execution rate is the infinitesimal counterpart of the cumulative schedule: if \(C(v)\) is the fraction executed by \(v\), then \(C(v)=\tfrac{1}{Q(F)}\int_0^v q(u)\,du\) with \(C(F)=1\). Thus different schedules \(C\) correspond to different rates \(q\) and hence to different impact trajectories.

With a linear propagator and power-law kernel \(G(t)\propto t^{-\rho}\), \(\rho=\tfrac12\) (close to the value implied by the schedules of hidden metaorders below), a constant-rate schedule \(q(u)=q_0\) yields \(\int_0^v q_0(v-u)^{-1/2}\,du\propto\sqrt{v}\): a square-root-like trajectory, the benchmark for visible metaorders. A schedule that trades faster near the beginning and the end generates instead a trajectory that is steep at both ends, the double inflection observed for hidden metaorders. Such a U-shaped schedule is the risk-neutral optimum: it minimizes the expected impact cost of executing a given quantity over a given horizon. Its form depends on the kernel, and this lets the data discriminate between the two standard specifications. With an exponentially decaying kernel, the resilience model of \citet{ObizhaevaWang2013}, the optimum consists of two discrete trades at \(\tau=0\) and \(\tau=1\) joined by execution at a constant rate, so the interior of the schedule is flat. With a power-law kernel \(G(t)\propto t^{-\rho}\) the optimal rate is instead \citep{gatheral2012transient}
\[
v(\tau)\;\propto\;\big[\tau(1-\tau)\big]^{-(1-\rho)/2},
\]
a smooth U that rises continuously towards both ends. Hidden metaorders follow the second form (Figure~\ref{fig:execution_schedule}): the rate has no flat interior, rising from \(0.66\) in the middle of the execution to \(0.73\) at \(\tau\approx0.2\) and \(0.93\) at \(\tau\approx0.07\), and the fitted exponent \(a=(1-\rho)/2=0.24\) implies \(\rho\approx0.5\) in the pooled sample (\(\rho\approx0.4\)--\(0.5\) for short and medium orders, \(\rho\approx0.7\) for long ones). The same U appears in each of the nine cells by tercile of duration and of participation rate: every cell executes between \(14\%\) and \(25\%\) of its volume in the first twentieth of the duration and in the last, against \(5\%\) at a constant rate, and the fitted exponent ranges from \(0.15\)--\(0.17\) for long orders to \(0.22\)--\(0.36\) for short and medium ones. The observed U is symmetric, as in the risk-neutral case. The benchmark thus links the schedules of Figure~\ref{fig:execution_schedule} to the trajectories of Figure~\ref{fig:rq3_binwise_raw_perp}. It is qualitative, not a structural calibration: it shows that the empirical shapes are consistent with transient-impact dynamics and that the difference between the two groups maps to the difference between near-uniform and U-shaped schedules.

One may suspect that the acceleration at the end comes from how we build hidden orders rather than from the trader's plan. An order ends at its last trade, and traders often stop because of the price: a stop-loss after an adverse move, or a profit target. Stopping of this kind decides when the order ends, but not how fast the trader trades before it: a trader who trades at a steady pace and stops when the price reaches a level leaves a flat schedule up to the last trade. The schedules show instead an acceleration that starts about a fifth of the duration before the end, as for a trader who knows when the order will be finished. Nor does the end depend on the price path. We split orders by whether the price moved in their direction or against them during the first three quarters of execution, and leave aside the first and last child trades. Between \(90\%\) and \(95\%\) of the duration the pace is the same in the two groups (\(1.01\) and \(1.02\) times uniform). The two groups differ as much in the first twentieth of the order, before any price move, as in the last (\(2.2\) against \(1.9\) in both). The end of a hidden metaorder is planned, not caused by the price.

\subsection{\texorpdfstring{Price Impact by Duration and Participation}{Price Impact by Duration and Participation}}
\label{app:levels}

Table~\ref{tab:levels-cells} breaks down by duration and participation the two levels behind the comparison of Section~\ref{subsec:ap-information}: the average price change at completion and the permanent price impact at \(\tau=4\). These are averages, without the controls of Table~\ref{tab:persistent-impact-discount}.

\begin{table}[H]
\centering
\small
{\begin{tabular}{llrrrrrr}
\toprule
 & & \multicolumn{2}{c}{Completion, \(\tau=1\)} & \multicolumn{2}{c}{Permanent, \(\tau=4\)} & \multicolumn{2}{c}{Orders (thousands)} \\
Duration & Participation & Hidden & Visible & Hidden & Visible & Hidden & Visible \\
\midrule
Short & \(0.008\)--\(0.12\) & 14.0 & 8.8 & 11.4 & 6.5 & 338 & 9 \\
Short & \(>0.12\) & 15.7 & 8.9 & 11.8 & 5.8 & 618 & 4 \\
Medium & \(0.008\)--\(0.12\) & 16.3 & 9.4 & 10.8 & 6.9 & 428 & 61 \\
Medium & \(>0.12\) & 19.1 & 12.9 & 13.2 & 10.0 & 401 & 45 \\
Long & \(0.008\)--\(0.12\) & 15.5 & 12.2 & 10.2 & 9.9 & 471 & 39 \\
Long & \(>0.12\) & 19.8 & 20.7 & 12.5 & 15.6 & 210 & 22 \\
\midrule
All & & 16.4 & 12.2 & 11.5 & 9.3 & 2,481 & 181 \\
\bottomrule
\end{tabular}
\caption{Average signed price change since the start of the metaorder, in basis points, at completion and at \(\tau=4\), by tercile of duration (cut-offs 3.7 and 23 minutes) and participation rate, on the sample of Table~\ref{tab:persistent-impact-discount} (participation at least \(0.0076\); \(1\%\) winsorized at each horizon).}
\label{tab:levels-cells}
}
\end{table}

\subsection{\texorpdfstring{The Bound on the Cost Shifted onto Nonannouncers}{The Bound on the Cost Shifted onto Nonannouncers}}
\label{app:bound}

This section derives the bound of Section~\ref{subsec:ap-nonannouncers}. In \citet{AdmatiPfleiderer1991}, let \(\gamma\) be the price response to liquidity demand when nobody announces. When uninformed traders announce, the price responds less to announced demand and more to unannounced demand: \(\gamma_A^{*}<\gamma<\gamma_N^{*}\). Let \(a\) and \(n\) be the variances of announced and unannounced liquidity demand, and \(s\) the variance of the noise in private signals. Their equilibrium price functions give \[\frac{\gamma_N^{*}-\gamma}{\gamma-\gamma_A^{*}}=\frac{s^{2}a}{s\,a+(s^{2}+s)\,n+1}<\frac{a}{n},\qquad\text{so that}\qquad \gamma_N^{*}-\gamma<\frac{a}{a+n}\,(\gamma_N^{*}-\gamma_A^{*}).\] The left-hand side of the second inequality is the extra price response paid by liquidity traders who do not announce; the right-hand side is the announced share of the variance of liquidity demand times the gap between the two responses. The share in the model is a ratio of variances of liquidity demand, which we do not observe. As an illustration, we use in its place the \(3.8\%\) of taker volume executed by visible metaorders, which is of the same order as the announced share of the variance of signed taker flow in our data, and the cost difference of ten basis points for the gap, which gives \(0.4\) basis points. If three quarters of unannounced volume is informed, the announced share of liquidity demand is \(3.8/(3.8+0.25\times 96.2)=14\%\) and the value is \(1.4\) basis points. These are orders of magnitude, not empirical bounds.

\subsection{\texorpdfstring{Other Samples and Specifications}{Other Samples and Specifications}}
\label{app:robustness}
Table~\ref{tab:robustness} repeats the estimates of Tables~\ref{tab:itmp-tradeoff-regression} and~\ref{tab:persistent-impact-discount} on other samples and specifications. In the first block neither discount changes by more than one basis point. With errors clustered by day and by market the two discounts of the main text remain significant. The last three rows use the visible metaorders whose on-chain status is observed, a part of the sample in which both discounts are larger (\(14.0\) and \(10.0\) basis points); they are to be compared with each other. Keeping only the visible metaorders that ran to completion, or controlling for their announced size and duration, changes the cost discount by about one basis point and the permanent-impact discount by about half a basis point or less.

\begin{table}[h!]
\centering
\footnotesize
\setlength{\tabcolsep}{4pt}
\begin{tabular}{lccccc}
\toprule
 & Cost discount & \multicolumn{2}{c}{Permanent-impact discount} & \multicolumn{2}{c}{Metaorders (thousands)} \\
\cmidrule(lr){3-4}\cmidrule(lr){5-6}
 & (bps) & \(\tau=1.5\) & \(\tau=4\) & Table~\ref{tab:itmp-tradeoff-regression} & Table~\ref{tab:persistent-impact-discount} \\
\midrule
Main text & \(-10.0\) \((0.4)\) & \(-4.80\) \((0.73)\) & \(-4.40\) \((0.97)\) & \(2{,}244\) / \(180\) & \(2{,}481\) / \(181\) \\
Transaction prices, same metaorders & \(-9.2\) \((0.4)\) & \(-5.53\) \((0.73)\) & \(-5.17\) \((0.97)\) & \(2{,}244\) / \(180\) & \(2{,}481\) / \(181\) \\
Volume and volatility over the previous 24 hours & \(-10.8\) \((0.4)\) & \(-5.02\) \((0.71)\) & \(-4.80\) \((0.95)\) & \(2{,}217\) / \(179\) & \(2{,}463\) / \(181\) \\
Without metaorders near liquidations & \(-9.9\) \((0.4)\) & \(-4.67\) \((0.79)\) & \(-4.24\) \((1.03)\) & \(2{,}241\) / \(180\) & \(2{,}478\) / \(181\) \\[3pt]
At least ten child orders in both groups & \(-10.8\) \((0.4)\) & \(-4.80\) \((0.76)\) & \(-4.50\) \((1.01)\) & \(2{,}199\) / \(164\) & \(2{,}481\) / \(164\) \\[3pt]
\multicolumn{6}{l}{\emph{Visible metaorders whose on-chain status is observed}} \\
\quad All visible metaorders & \(-14.0\) \((0.6)\) & \(-10.02\) \((0.87)\) & \(-10.50\) \((1.35)\) & \(838\) / \(67\) & \(973\) / \(72\) \\
\quad Completed visible metaorders & \(-15.2\) \((0.6)\) & \(-10.33\) \((0.84)\) & \(-10.02\) \((1.35)\) & \(792\) / \(48\) & \(973\) / \(54\) \\
\quad Announced size and duration as controls & \(-14.5\) \((0.6)\) & \(-10.06\) \((0.87)\) & \(-9.91\) \((1.36)\) & \(837\) / \(67\) & \(973\) / \(72\) \\
\bottomrule
\end{tabular}
\caption{The discounts of Tables~\ref{tab:itmp-tradeoff-regression} and~\ref{tab:persistent-impact-discount} on other samples and specifications, midquotes except in the second row. Cost discount: coefficient on visible metaorders in Table~\ref{tab:itmp-tradeoff-regression}, in basis points at the median volatility of each sample. Permanent-impact discount: coefficient of Table~\ref{tab:persistent-impact-discount}, in basis points. Previous 24 hours: the volume and volatility of the normalizations are measured before the start instead of after it. Near liquidations: hidden metaorders with a liquidation order of the same address and market inside them or within 30 minutes of their start or end. Controls: the logarithms of planned over executed size and of planned over realized duration. Metaorders: hidden / visible. Standard errors clustered by day in parentheses.}
\label{tab:robustness}
\end{table}

\subsection{\texorpdfstring{Predictability of Order Arrivals}{Predictability of Order Arrivals}}
\label{app:arrivals}
This section reports the prediction exercise of Section~\ref{sec:robustness}. We predict, for each market and five-minute interval, whether a visible metaorder starts, in six monthly out-of-sample tests from October 2025 to March 2026, each with a model fitted on all earlier months. A baseline with market and hour-of-day effects has an area under the precision--recall curve (PR-AUC) of \(0.31\), against the base rate of \(2.0\%\) for a model with no information; adding the state of the market before the interval raises it to \(0.42\). Almost all of the gain comes from recent activity, the visible and hidden metaorders started in the market over the previous 24 hours: once activity and recent taker flow are included, removing spread, depth and imbalance leaves the PR-AUC unchanged (difference \(-0.0001\), \(95\%\) interval \([-0.0002,\,0.0001]\)). For starts of hidden metaorders the same variables raise it from \(0.43\) to \(0.55\).

\section{\texorpdfstring{Appendix: Trader Composition}{Appendix: Trader Composition}}
    \label{app:traders}
    Trading on Hyperliquid is concentrated, and the regressions of the main text give the same weight to every metaorder. The \(1\%\) of addresses with the most metaorders (248 addresses in the sample of Table~\ref{tab:persistent-impact-discount}) send \(86\%\) of the hidden and \(26\%\) of the visible metaorders, and the two strategies of the protocol's market-making vault (HLP) alone send \(16\%\) of the hidden ones. Table~\ref{tab:traders} shows how the two discounts depend on this composition.

\begin{table}[h!]
\centering
\footnotesize
\begin{tabular}{lccc}
\toprule
 & Cost discount & \multicolumn{2}{c}{Permanent-impact discount} \\
\cmidrule(lr){3-4}
 & (bps) & \(\tau=1.5\) & \(\tau=4\) \\
\midrule
All metaorders, one weight per metaorder (main text) & \(-10.0\) \((0.4)\) \([4.1]\) & \(-4.80\) \((0.73)\) \([2.52]\) & \(-4.40\) \((0.97)\) \([2.52]\) \\
One weight per address & \(-7.0\) \((1.0)\) \([1.1]\) & \(-4.12\) \((1.87)\) \([1.88]\) & \(-3.62\) \((3.06)\) \([3.07]\) \\
Same address (address fixed effects) & \(-4.1\) \((1.1)\) \([3.1]\) & \(-6.89\) \((1.57)\) \([2.92]\) & \(-8.02\) \((2.55)\) \([3.23]\) \\
Same entity (entity fixed effects) & \(-9.8\) \((1.0)\) & \(-6.73\) \((1.44)\) & \(-6.97\) \((2.41)\) \\[3pt]
\multicolumn{4}{l}{\emph{Without the \(1\%\) of addresses with the most metaorders}} \\
\quad One weight per metaorder & \(-2.2\) \((0.6)\) \([1.3]\) & \(-1.87\) \((0.65)\) \([1.08]\) & \(-2.12\) \((1.02)\) \([1.31]\) \\
\quad One weight per address & \(-6.2\) \((1.3)\) \([1.3]\) & \(-4.05\) \((1.90)\) \([1.91]\) & \(-3.51\) \((3.09)\) \([3.10]\) \\
\quad Same address & \(-6.4\) \((1.0)\) \([1.9]\) & \(-11.08\) \((1.59)\) \([2.31]\) & \(-11.97\) \((2.85)\) \([3.24]\) \\[3pt]
Only the \(1\%\) of addresses with the most metaorders & \(-11.9\) \((0.6)\) \([4.8]\) & \(-6.77\) \((1.05)\) \([3.49]\) & \(-5.72\) \((1.49)\) \([3.47]\) \\
Without the \(1\%\) of addresses with the largest value & \(-10.5\) \((0.8)\) & \(-5.39\) \((0.62)\) & \(-4.87\) \((1.02)\) \\
Without vault addresses & \(-18.2\) \((0.5)\) & \(-8.79\) \((0.72)\) & \(-8.28\) \((0.99)\) \\
Without vault addresses and liquidations & \(-18.1\) \((0.5)\) & \(-8.72\) \((0.73)\) & \(-8.21\) \((1.00)\) \\
\bottomrule
\end{tabular}
\caption{The discounts of Tables~\ref{tab:itmp-tradeoff-regression} and~\ref{tab:persistent-impact-discount} by trader composition, midquotes; columns as in Table~\ref{tab:robustness}. One weight per address: every address has total weight one among visible and among hidden metaorders. Same address: address fixed effects, identified by the addresses with metaorders of both groups. Same entity: fixed effects of the entity, which groups the sub-accounts of a master account, on the sample of the main text. Standard errors clustered by day in parentheses and by day and address in brackets.}
\label{tab:traders}
\end{table}

With one weight per address the cost discount is \(7.0\) basis points instead of \(10.0\), and its standard error clustered by day and address falls from \(4.1\) to \(1.1\); the permanent-impact discount is \(4.1\) basis points at \(\tau=1.5\) (s.e.\ \(1.9\)) and is not significant at \(\tau=4\). Addresses outside the most active \(1\%\) have about the same discounts when each counts once, and the same address among them pays \(6.4\) basis points less when it announces and leaves \(11\) to \(12\) basis points less permanent price impact. With one weight per metaorder, instead, the discount outside the most active addresses is small (\(2.2\) basis points): the average of the main text is set by a few hundred active accounts, which alone give \(11.9\). Without vault addresses both discounts almost double. With errors clustered by day and address, the cost discount of the main text remains significant (\(t=-2.4\)) and the permanent-impact discount is borderline (\(t=-1.9\) at \(\tau=1.5\) and \(-1.7\) at \(\tau=4\)).

The same concentration matters for the tests of Hypothesis~\ref{hyp:nonannouncer}. In the sample of hidden metaorders of Section~\ref{subsec:ap-nonannouncers}, the \(1\%\) of addresses with the largest traded value (206 addresses) execute \(85\%\) of the value and \(67\%\) of the metaorders. Table~\ref{tab:traders-h3} repeats the tests of Hypothesis~\ref{hyp:nonannouncer} without the most active addresses, ranked in two ways: the \(1\%\) with the largest traded value, and the \(1\%\) with the most metaorders, as in Table~\ref{tab:traders}. Only their own metaorders are dropped; the flow running alongside each metaorder is still measured on all addresses.

    \begin{table}[h!]
    \centering
    \small
    \begin{tabular}{lccc}
    \toprule
    & & \multicolumn{2}{c}{Without the top \(1\%\) of addresses} \\
    \cmidrule(lr){3-4}
    & All addresses & by value & by number \\
    \midrule
    Same direction, visible minus hidden & \(+0.76\) \((0.69)\) & \(+5.78\) \((1.28)\) & \(+4.33\) \((1.50)\) \\
    Opposite direction, visible minus hidden & \(+0.93\) \((0.73)\) & \(-8.18\) \((1.49)\) & \(-9.17\) \((1.66)\) \\
    Sum, independent of the side & \(+1.69\) \((0.81)\) & \(-2.40\) \((1.63)\) & \(-4.85\) \((2.22)\) \\
    Hidden metaorders & \(2{,}134{,}204\) & \(720{,}724\) & \(347{,}904\) \\[3pt]
    Adoption, per \(10\) points of announced share & \(+0.24\) \((0.89)\) & \(-1.28\) \((1.73)\) & \(+1.16\) \((1.96)\) \\
    Hidden metaorders & \(2{,}302{,}449\) & \(741{,}951\) & \(348{,}175\) \\
    \bottomrule
    \end{tabular}
    \caption{Tests of Hypothesis~\ref{hyp:nonannouncer} without the most active addresses. Cost of hidden metaorders at completion, in basis points, midquotes. First three rows: coefficients of the regression of Section~\ref{subsec:ap-nonannouncers} on the shares of the four groups of flow, per unit share, differences between visible and hidden flow on each side; standard errors clustered by day. Adoption: coefficient on the market-month share of taker volume executed by visible metaorders; standard errors clustered by market.}
    \label{tab:traders-h3}
    \end{table}

    Without the most active addresses, visible flow raises the cost of hidden metaorders more than hidden flow does when it trades in their direction, and lowers it more when it trades against them, so the part that does not depend on the side is negative, significantly so when addresses are ranked by number. Visible flow thus moves prices more per unit of share than hidden flow. This is price pressure, not the cost of announcement predicted by \citet{AdmatiPfleiderer1991}, which would raise the cost of hidden metaorders whatever the side of the visible flow. With all addresses the part that does not depend on the side is positive (\(1.7\), s.e.\ \(0.8\); \(1.0\) with errors clustered by day and market), but it comes from the most active addresses and does not survive their exclusion. The same holds when visible flow is measured by the size announced and not yet executed: the sum is \(5.1\) (s.e.\ \(1.3\)) with all addresses, and \(-0.4\) (s.e.\ \(2.7\)) and \(0.3\) (s.e.\ \(3.0\)) without the most active addresses ranked by value and by number. The adoption test remains insignificant. 

    \section{\texorpdfstring{Appendix: Alternative Constructions of Hidden Metaorders}{Appendix: Alternative Constructions of Hidden Metaorders}}
\label{app:construction}

\subsection{\texorpdfstring{Other Pauses}{Other Pauses}}
\label{app:gaps}
The main text ends a hidden metaorder at a pause longer than 30 minutes. Table~\ref{tab:gaps} rebuilds the hidden metaorders with pauses of 5, 15 and 60 minutes, with the same filters (at least ten child orders, at most 24 hours) and the same specifications as Tables~\ref{tab:itmp-tradeoff-regression} and~\ref{tab:persistent-impact-discount}; the 30-minute midquote rows are those of the main text. Every sign and significance level is unchanged, with midquotes and with transaction prices. The cost discount is about the same with 5- and 15-minute pauses and larger with longer ones; the permanent-impact discount grows with the pause. 

\begin{table}[H]
\centering
\small
\begin{tabular}{lcccc}
\toprule
Pause that ends a metaorder & Hidden metaorders & Cost discount \(\hat\theta\) & \multicolumn{2}{c}{Permanent-impact discount (bps)} \\
\cmidrule(lr){4-5}
 & & & \(\tau=1.5\) & \(\tau=4\) \\
\midrule
\multicolumn{5}{l}{\emph{Midquotes}} \\
5 minutes & \(3{,}822{,}299\) & \(-0.0173\) \((0.0007)\) & \(-3.26\) \((0.62)\) & \(-3.12\) \((0.84)\) \\
15 minutes & \(4{,}181{,}883\) & \(-0.0171\) \((0.0008)\) & \(-3.63\) \((0.68)\) & \(-3.44\) \((0.92)\) \\
30 minutes (main text) & \(4{,}271{,}097\) & \(-0.0202\) \((0.0009)\) & \(-4.80\) \((0.73)\) & \(-4.40\) \((0.97)\) \\
60 minutes & \(4{,}320{,}027\) & \(-0.0225\) \((0.0009)\) & \(-5.63\) \((0.76)\) & \(-5.13\) \((1.03)\) \\[3pt]
\multicolumn{5}{l}{\emph{Transaction prices, all metaorders}} \\
5 minutes & & \(-0.0155\) \((0.0006)\) & \(-3.54\) \((0.61)\) & \(-3.66\) \((0.82)\) \\
15 minutes & & \(-0.0150\) \((0.0008)\) & \(-4.00\) \((0.67)\) & \(-4.03\) \((0.89)\) \\
30 minutes (main text) & & \(-0.0178\) \((0.0009)\) & \(-5.24\) \((0.70)\) & \(-5.08\) \((0.93)\) \\
60 minutes & & \(-0.0199\) \((0.0009)\) & \(-6.10\) \((0.73)\) & \(-5.86\) \((0.97)\) \\
\bottomrule
\end{tabular}
\caption{Tables~\ref{tab:itmp-tradeoff-regression} and~\ref{tab:persistent-impact-discount} with hidden metaorders rebuilt with other pauses. Hidden metaorders: those with at least ten child orders and at most 24 hours. Cost discount: coefficient on visible metaorders in Table~\ref{tab:itmp-tradeoff-regression}, in units of daily volatility; permanent-impact discount: coefficient of Table~\ref{tab:persistent-impact-discount}. Same specifications as in the main text; standard errors clustered by day. Transaction-price rows use all metaorders, midquote rows those whose midquotes are observed.}
\label{tab:gaps}
\end{table}

\subsection{Hidden Metaorders from All Fills}
\label{app:fmo}

The hidden metaorders of the main text are built from market orders only, so they could include very active liquidity providers who repeatedly cross the spread. Among reconstructed metaorders with at least ten child orders, \(40.3\%\) have a same-direction maker fill of the same address during execution, but this passive volume is highly concentrated: the top 40 addresses by passive maker-fill volume account for about \(65\%\) of same-direction passive USD, and they display both same- and opposite-direction maker fills, consistent with market-making rather than diffuse passive metaorder execution.  We therefore keep the definition based on market orders and do not add passive fills to size or participation.

A second check changes the definition itself. \citet{moro2009market} reconstruct hidden orders from all the trades of a participant and keep those executed mostly with market orders. We take from them the use of all fills and the criterion of market-order execution. For each address and market we form runs of consecutive same-direction fills, from both market and limit orders, with the same rules as in Section~\ref{sec:setting} (30-minute gap, at least ten orders, duration up to 24 hours), and compute the fraction \(f_{mo}\) of the run's notional executed with market orders. Of the resulting runs, \(3.13\) million have \(f_{mo}>0.8\), against \(4.27\) million hidden metaorders in the main text. In \(39\%\) of these runs at least one fill comes from a limit order, but limit orders execute only \(2.8\%\) of their notional, and about three fifths of the hidden metaorders of the main text belong to an all-fills run with \(f_{mo}>0.8\). Runs with \(f_{mo}<0.2\), which are liquidity provision rather than liquidity demand, have negative impact and are not used. Prices, volatility and participation are measured as in the main text.

Table~\ref{tab:fmo} repeats the tests of Hypotheses~\ref{hyp:announcer} and~\ref{hyp:information} with these runs. The permanent-impact discount keeps its sign and significance. With midquotes it is smaller than in the main text at \(\tau=1.5\) (\(-3.56\) against \(-4.80\) basis points) and about the same at \(\tau=4\) (\(-4.08\) against \(-4.40\)); with transaction prices it is about the same at \(\tau=1.5\) and larger at \(\tau=4\). The cost discount is the same as in the main text with midquotes (\(-0.0202\)) and slightly larger with transaction prices (\(-0.0195\) against \(-0.0178\)). The share of impact that remains after completion differs between the two constructions: computed as in Figure~\ref{fig:impact_decay}, at \(\tau=4\) hidden metaorders built from all fills keep \(84.3\%\) of their impact at completion, against \(70.6\%\) for those of the main text. The difference between visible and hidden metaorders in this share is not significant with either construction (\(-11.9\) percentage points, s.e.\ \(7.4\), against \(+3.0\), s.e.\ \(7.0\)), while the gap in basis points is significant with both (\(-4.41\), s.e.\ \(0.99\), against \(-3.06\), s.e.\ \(0.97\)).

\begin{table}[H]
\centering
\small
\begin{tabular}{lcccc}
\toprule
Hidden metaorders & Number & Cost discount \(\hat\theta\) & \multicolumn{2}{c}{Permanent-impact discount (bps)} \\
\cmidrule(lr){4-5}
 & & & \(\tau=1.5\) & \(\tau=4\) \\
\midrule
\multicolumn{5}{l}{\emph{Midquotes}} \\
Market orders (main text) & \(4{,}271{,}097\) & \(-0.0202\) \((0.0009)\) & \(-4.80\) \((0.73)\) & \(-4.40\) \((0.97)\) \\
All fills, \(f_{mo}>0.8\) & \(3{,}131{,}916\) & \(-0.0202\) \((0.0008)\) & \(-3.56\) \((0.68)\) & \(-4.08\) \((0.98)\) \\[3pt]
\multicolumn{5}{l}{\emph{Transaction prices, all metaorders}} \\
Market orders (main text) & & \(-0.0178\) \((0.0009)\) & \(-5.24\) \((0.70)\) & \(-5.08\) \((0.93)\) \\
All fills, \(f_{mo}>0.8\) & & \(-0.0195\) \((0.0008)\) & \(-5.00\) \((0.65)\) & \(-5.85\) \((0.93)\) \\
\bottomrule
\end{tabular}
\caption{Tables~\ref{tab:itmp-tradeoff-regression} and~\ref{tab:persistent-impact-discount} with hidden metaorders rebuilt from all fills of an address, keeping runs with \(f_{mo}>0.8\). Columns, specifications and standard errors as in Table~\ref{tab:gaps}; the sample of visible metaorders is unchanged. Transaction-price rows use all metaorders, midquote rows those whose midquotes are observed.}
\label{tab:fmo}
\end{table}

\section{\texorpdfstring{Appendix: Interrupted Visible Metaorders}{Appendix: Interrupted Visible Metaorders}}
\label{app:interrupted}
A visible metaorder can stop before its announced size is filled, so that part of the announcement turns out untrue. Its end is as public as its start: a cancellation is a transaction recorded on chain, like the order itself, and a stop for insufficient margin appears in the on-chain status of the order, so other traders can see that the rest of the announced size will not be executed. About a third of the TWAPs that trade are cancelled by the trader and one in seven is stopped for insufficient margin (Section~\ref{sec:regimes-definition}). Cancelled programs stop after about a fifth of their planned size, and cancelling is the habit of a few addresses: those with at least ten cancellations account for \(91\%\) of them. Traders cancel after a small move against them (\(2.5\) basis points more than completed TWAPs at the same point of execution, s.e.\ \(0.6\)), while margin stops follow a sharp move against the trader's position. Interruption does not undo the results: cancelled TWAPs keep about two thirds of the cost advantage of completed ones over hidden metaorders, and their permanent impact at \(\tau=4\) is the same as that of completed TWAPs (difference \(0.3\) basis points, s.e.\ \(2.2\)). The market reacts to a cancellation, but little and briefly. Measured from the cancellation and relative to random times of the same market and day, the midquote retraces by \(0.45\) basis points in the first minute (s.e.\ \(0.10\)), against \(0.02\) (s.e.\ \(0.07\)) after the last child of a completed TWAP. The retracement is larger when the cancelled remainder is large (\(1.6\) basis points, s.e.\ \(0.3\), when it exceeds \(0.1\%\) of the market's daily volume) and is no longer detectable after fifteen minutes.

\end{document}